\documentclass[a4paper,8pt, aps, prl, reprint, superscriptaddress, longbibliography]{revtex4-1}  % put 'reprint' for two-column, 'preprint' for one-column
\usepackage[utf8]{inputenc}
\usepackage{dcolumn} % Align table columns on decimal point
\usepackage{bm} % bold math
\usepackage{amsmath} % provides various features for displayed equations
\usepackage{amssymb}
\usepackage{xr}
\usepackage{graphicx}
\usepackage{subfigure}
\usepackage{gensymb} %to write a ° (degree) symbol
\usepackage{natbib}
\usepackage{url}
\usepackage{textcase}
\usepackage{amsfonts}
\usepackage{lineno}
\usepackage{color}

\usepackage{fancyhdr} %Headers and footers
\usepackage{setspace} %Line settings
\usepackage{geometry}
\begin{document}

%\linenumbers

\title{The importance of temperature-dependent collision frequency in PIC simulation on nanometric density evolution of highly-collisional strongly-coupled dense plasmas}

\author{Mohammadreza Banjafar}
\email{mohammadreza.banjafar@xfel.eu}
\affiliation{European XFEL, Holzkoppel 4, 22869, Schenefeld, Germany}

\author{Lisa Randolph}
\affiliation{Department Physik, Universit\"{a}t Siegen, Walter-Flex-Str. 3,  57072, Siegen, Germany}

\author{Lingen Huang}
\affiliation{Helmholtz-Zentrum Dresden-Rossendorf, Bautzner Landstraße 400, 01328, Dresden, Germany}

\author{S.V. Rahul}
\affiliation{European XFEL, Holzkoppel 4, 22869, Schenefeld, Germany}

\author{Thomas R. Preston}
\affiliation{European XFEL, Holzkoppel 4, 22869, Schenefeld, Germany}

\author{Toshinori Yabuuchi}
\affiliation{Japan Synchrotron Radiation Research Institute (JASRI), Sayo, Hyogo, 679-5198, Japan}
\affiliation{RIKEN SPring-8 Center, Sayo, Hyogo 679-5148, Japan}

\author{Mikako Makita}
\affiliation{European XFEL, Holzkoppel 4, 22869, Schenefeld, Germany}

\author{Nicholas P. Dover}
\affiliation{QST-Kansai, KPSI,8-1-7 Umemi-dai, Kizugawa-city, Kyoto, 619-0215, Japan }

\author{Sebastian G\"{o}de}
\affiliation{European XFEL, Holzkoppel 4, 22869, Schenefeld, Germany}

\author{Akira Kon}
\affiliation{QST-Kansai, KPSI,8-1-7 Umemi-dai, Kizugawa-city, Kyoto, 619-0215, Japan }

\author{James K. Koga}
\affiliation{QST-Kansai, KPSI,8-1-7 Umemi-dai, Kizugawa-city, Kyoto, 619-0215, Japan }

\author{Mamiko Nishiuchi}
\affiliation{QST-Kansai, KPSI,8-1-7 Umemi-dai, Kizugawa-city, Kyoto, 619-0215, Japan }

\author{Michael Paulus}
\affiliation{Technical University Dortmund, Otto-Hahn Straße 4, 44227 Dortmund, Germany}

\author{Christian R\"{o}del}
\affiliation{Technical University Darmstadt, Schlossgartenstraße 9, 64289 Darmstadt, Germany}

\author{Michael Bussmann}
\affiliation{Helmholtz-Zentrum Dresden-Rossendorf, Bautzner Landstraße 400, 01328, Dresden, Germany}

\author{Thomas E. Cowan}
\affiliation{Technical University Dresden, 01069 Dresden, Germany}
\affiliation{Helmholtz-Zentrum Dresden-Rossendorf, Bautzner Landstraße 400, 01328, Dresden, Germany}

\author{Christian Gutt}
\affiliation{Department Physik, Universit\"{a}t Siegen, Walter-Flex-Str. 3,  57072, Siegen, Germany}

\author{Adrian P. Mancuso}
\affiliation{European XFEL, Holzkoppel 4, 22869, Schenefeld, Germany}
\affiliation{Department of Chemistry and Physics, La Trobe Institute for Molecular Science, La Trobe University, Melbourne, Victoria 3086, Australia}

\author{Thomas Kluge}
\affiliation{Helmholtz-Zentrum Dresden-Rossendorf, Bautzner Landstraße 400, 01328, Dresden, Germany}

\author{Motoaki Nakatsutsumi}
\affiliation{European XFEL, Holzkoppel 4, 22869, Schenefeld, Germany}
\affiliation{Open and Transdisciplinary Research Institute, Osaka University, Suita, Osaka 565-0087, Japan}

\date{\today}

\begin{abstract}
    \noindent Particle-in-Cell (PIC) method is a powerful plasma simulation tool for investigating high-intensity femtosecond laser-matter interaction. However, its simulation capability at high-density plasmas around the Fermi temperature is considered to be inadequate due, among others, to the necessity of implementing atomic-scale collisions. Here, we performed a one-dimensional with three-velocity space (1D3V) PIC simulation that features the realistic collision frequency around the Fermi temperature and atomic-scale cell size. The results are compared with state-of-the-art experimental results as well as with hydrodynamic simulation. We found that the PIC simulation is capable of simulating the nanoscale dynamics of solid-density plasmas around the Fermi temperature up to $\sim$2~ps driven by a laser pulse at the moderate intensity of $10^{14-15}$~$\mathrm{W/cm^{2}}$, by comparing with the state-of-the-art experimental results. The reliability of the simulation can be further improved in the future by implementing multi-dimensional kinetics and radiation transport.
    
\end{abstract}

\maketitle 

\section{INTRODUCTION}
\noindent

    The interaction of sub-picosecond, high-intensity short-pulse lasers with solids at moderate laser intensities ($10^{14-15}$~$\mathrm{W/cm^{2}}$) generates strongly-coupled dense plasmas with a few eV electron temperatures, i.e., the electron thermal energy is comparable to the Fermi energy~\cite{lee_baldis2002}. Under such conditions, atomic-scale collisions and quantum effects such as electron degeneracy and ion correlations play an essential role in electron dynamics~\cite{rightley2021, moldabekov2015}. Such plasmas can be found inside planets~\cite{Remington2000}, inertial confinement fusion (ICF) plasmas ~\cite{Glenzer2010,atzeni2004}, femtosecond laser ablation for micro-processing~\cite{Bonse2015,Nolte1997} and early stage of ultra-high intensity laser-solid interaction at election relativistic regime. 
    
    The electromagnetic Particle-in-Cell (PIC) simulation method is widely used to model high-intensity laser-matter interactions~\cite{SENTOKU2008, Antici2008, Kluge2016, macchi2018, Huang2016}. PIC solves Maxwell's equations on a grid system and applies relativistic equations to all particles, each representing an average of many real particles~\cite{birdsall2018}. This allows the evaluation of particle transport and kinetic effects in plasmas. To account for binary interactions between particles, a Monte Carlo model is used in the PIC algorithm~\cite{TA77, SENTOKU2008}. Recently, an experimental platform to benchmark the collision model in the PIC simulations at the regime of sub-relativistic laser intensity of $\sim10^{18}$~$\mathrm{W/cm^{2}}$ has been developed ~\cite{yang2023}. However, the reliability of PIC simulations in highly collisional degenerate regimes was always in doubt. Challenges include uncertainties in the electron collision frequency models, unreliability of the classical approximation for the momentum transport cross-section, i.e., Coulomb logarithm~\cite{rightley2021kinetic, starrett2018coulomb, filippov2018coulomb}, and the necessity to consider quantum effects and large-angle scattering and macro-fields~\cite{daligault2016quantum}. Among these challenges, the uncertainty of the electron collision frequency ($\nu$) is at most important, which governs the laser energy absorption~\cite{Price1995, Eidmann00}, electron-to-ion energy transfer ~\cite{VORBERGER2014}, heat diffusion~\cite{eliezer}, and subsequent surface expansion and ablation. 

    The electron collision frequency around the Fermi temperature ($T_F$) and its effects on short-pulse laser absorption and electron thermalization has been studied experimentally~\cite{fourment2014, Fisher01} and theoretically~\cite{Fisher01, mueller2013, meyer2019, lugovskoy1999, petrov2013, pineau2020modeling}. Fisher et al. \cite{Fisher01} demonstrated that it is essential to account for both electron-electron and electron-phonon collisions in the determination of the electron momentum relaxation rate in the interaction of femtosecond laser pulses with an intensity of $I\leqslant10^{15}$~$\mathrm{W/cm^2}$ in metallic solids~\cite{Fisher01}. Mueller and Rethfeld \cite{mueller2013} derived Boltzmann collision integrals to study electron excitation, thermalization, and electron-phonon coupling under non-equilibrium electron distribution. Petrov et al. \cite{petrov2013} calculated the thermal conductivity and electron-ion heat transfer coefficients for the interaction of short-pulse lasers and solids in the degenerate plasma regime~\cite{petrov2013}. Meyer-ter-Vehn and Ramis derived a global analytical expression for the effective collision frequency $\nu(T_e, E_{ph}, n_e)$~\cite{meyer2019}. Here, $T_e$, $n_e$, and $E_{ph}$ are electron temperature, electron density, and absorbed photon energy, respectively. Furthermore, Pineau et al. \cite{pineau2020modeling} modeled the effective electron collision frequency over the entire temperature range of solid-to-plasma transitions in ICF experiments.
    
    In this article, we implemented an improved temperature-dependent binary Coulomb collision model around the Fermi temperature ($T_F$) in one-dimensional with three-dimensional velocity space (1D3V) PICLS code~\cite{SENTOKU2008} to study nano-scale density dynamics and heatwave propagation inside femtosecond (fs) laser-excited solid-density plasmas. Angstrom-scale cell size is also implemented to resolve atomic-scale electron collisions. The results are then quantitatively compared with the recent experimental results. We also discuss the comparison with 1D~MULTI-fs hydrodynamic code~\cite{Eidmann00, Ramis2012}. 

    \section{Temperature-dependent electron collision frequency at low temperatures}
\begin{figure}
         \centering
         \includegraphics[width=1\linewidth]{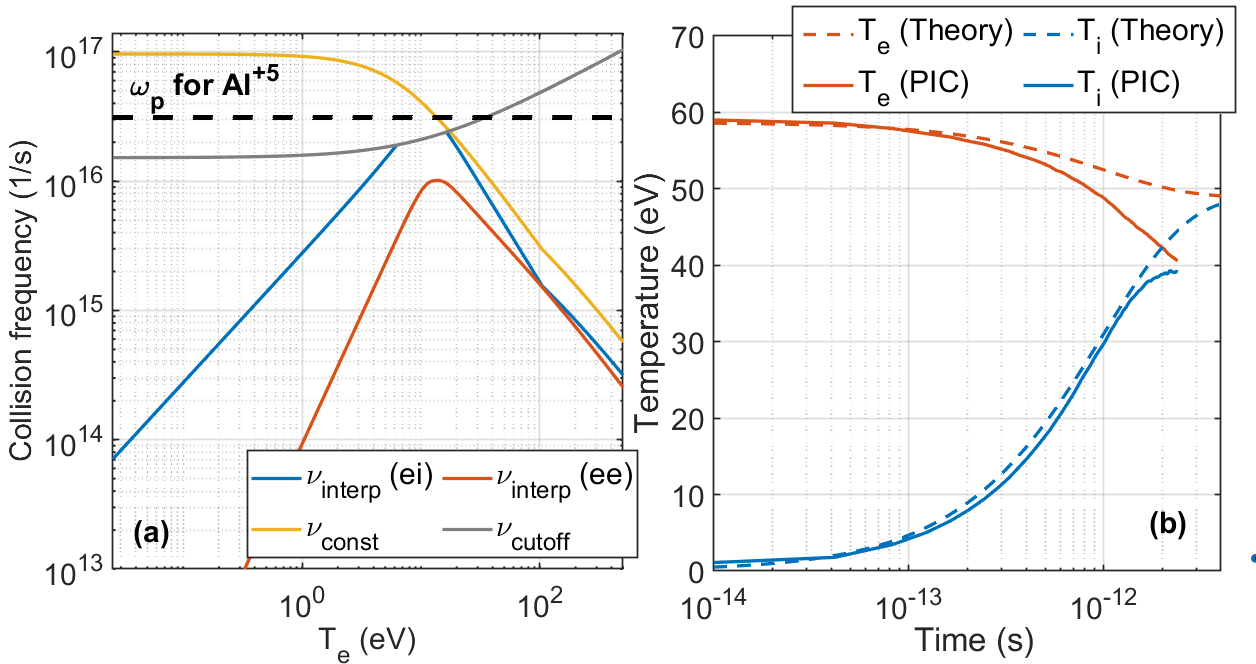}
         \vspace*{-6mm}
         \caption{(a) Interpolated electron collision frequency $\nu_{interp}$ vs. the electron temperature implemented in 1D~PICLS simulations for pre-ionized aluminum $\mathrm{Al^{5+}}$. The solid blue and red lines are the electron-ion ($\nu_{ei}^{-1} = \nu_{ei_{plasma}}^{-1} + \nu_{el-ph}^{-1}$) and electron-electron ($\nu_{ee}^{-1} = \nu_{ee_{plasma}}^{-1} + \nu_{ee_{metal}}^{-1}$) collision frequencies implemented in the simulation. The electron-phonon collision frequency ($\nu_{el-ph}$) assumes $T_i = T_e$. The solid yellow line is the collision frequency model used previously in PICLS with the constant value at low temperatures below $T_F$ ($\nu_{const}$). The cutoff collision frequency ($\nu_{cutoff}$) and the plasma frequency ($\omega_p$) for $\mathrm{Al^{5+}}$ are shown by the solid grey and dashed black lines, respectively. Here, the $\nu_{cutoff}$ is where the electron mean free path equals the ion sphere radius. (b) Electron (red) and ion (blue) thermalization dynamics extracted from PICLS using $\nu_{interp}$ (solid lines) and the theoretical values using the Eq.~\ref{eq:thermalization} (dashed lines). The initial electron and ion temperatures were set to 58 and 0.5~eV, respectively. The ionization was kept constant to $\mathrm{Al^{5+}}$.}
         \label{fig:collFreqPlots}
     \end{figure}

    Binary Coulomb collision between weighted particles has been implemented in PICLS~\cite{SENTOKU2008}, based on the Takizuka and Abe model~\cite{TA77}. Initially, particles are grouped in each cell, and then the code randomly determines which pairs of particles are likely to collide. In the next step, the alterations in the velocity of particles caused by the binary collision during the time interval $\Delta t$ are computed and applied to advance the particle velocities. In an ideal plasma (high temperature and low density), the collision frequency of the small-angle scattering between two sample particles $\alpha$ and $\beta$ can be obtained using the Spitzer formula as ~\cite{eliezer, Spitzer};
  
    \begin{equation}
    \nu_{plasma} = \frac{4 \pi (q_{\alpha}q_{\beta})^2 n_l \ln \Lambda}{p_{rel}^2 v_{rel}},
    \label{eq:binaryCollFrequency}
    \end{equation}
    
    \noindent where $q_{\alpha}$ and $q_{\beta}$ are the charge of the particles, $p_{rel} = m_{\alpha \beta} v_{rel}$ is the relative momentum between two particles, $m_{\alpha \beta}$ is the reduced mass, $v_{rel}$ is the relative velocity, $n_l$ is the lower density among $n_{\alpha}$ and $n_{\beta}$, and $\ln \Lambda = \ln{(\lambda_{De}/\lambda_{dB})}$ is the Coulomb logarithm. 
    Here, $\lambda_{D_e} = (k_B T_e/4\pi n_e e^2)^{1/2}$ is the Debye length and $\lambda_{dB}=h/2m_e v_e$ is de-Broglie's wavelength, where $k_B$ is the Boltzmann constant, $e$ and $v_e$ are the electron charge and velocity, respectively, and $h$ is the Planck's constant.
    Considering the thermal velocity of electrons $v_{th} = \sqrt{3k_BT_e/2m_e}$, where $m_e$ is the electron mass, the denominator in Eq.~\ref{eq:binaryCollFrequency} is proportional to $T_e^{3/2}$. 
    Therefore, the Spitzer collision frequency diverges at low $T_e$. To avoid this, the Fermi temperature $T_F$ was considered as the threshold temperature for the degenerate plasma regime, below which the collision frequency was set constant as $\nu_{const} = 4m_eZe^4\ln{\Lambda}/3\pi\hbar^3$~\cite{SENTOKU2008} (see the orange line in Fig~\ref{fig:collFreqPlots}), where $Z$ is the ionization state and $\hbar$ is the reduced Planck constant.  \\
    
    To improve the model at low-temperature and high-density plasmas around and below $T_F$, we implemented the temperature-dependent collision frequency at $T \leq T_F$ into PICLS based on the model using an \textit{ad hoc} interpolation as  Ref.~\cite{Eidmann00, Ramis2012}:
    
    \begin{equation}
       \nu_{interp}^{-1} = \nu_{plasma}^{-1} + \nu_{metal}^{-1}.
       \label{coll_interp}
    \end{equation}
    
    \noindent Eq.~\ref{coll_interp} smoothly bridges the collision frequency at high $T$ ideal plasmas ($\nu_{plasma}$) and cold metallic solids $\nu_{metal}$ ($T_e << T_F$). The $\nu_{metal}$ is the summation of the electron-electron $\nu_{ee}$ and electron-phonon $\nu_{el-ph}$ contribution:
    
    \begin{equation}
        \begin{split}
            & \nu_{ee_{metal}}(T_e) = \frac{E_F}{\hbar} \bigg(\frac{T_e}{T_F}\bigg)^{2},\\
            & \nu_{el-ph}(T_i) = K_{s}\frac{e^2 k_B T_i}{\hbar v_F},
            \end{split}
        \label{eq:nu_electron-phonon}
    \end{equation}
    
    \noindent where $E_F = k_B T_F = m_e v_F^2 / 2$ is the Fermi energy, $v_F = \hbar (3 \pi^2 n_e)^{1/3} / m_e$ is the Fermi velocity, $T_i$ is the ionic temperature, and $K_s$ is an empirical parameter which is adjusted using experimental observations~\cite{Eidmann00,Ramis2012}.
    At $T_e >> T_F$, the Spitzer formula (Eq.~\ref{eq:binaryCollFrequency}) includes all types of collisions; electron-electron ($\nu_{ee_{plasma}}$), electron-ion ($\nu_{ei_{plasma}}$), and ion-ion ($\nu_{ii_{plasma}}$). The Coulomb logarithm in Eq.~\ref{eq:binaryCollFrequency} is also modified to be~\cite{Ramis2012}
    
    \begin{equation}
        \ln{\Lambda} = \frac{1}{2} \ln{[1 + \max(1,{b_{max}}/{b_{min}})]}
        \label{lnLambda}
    \end{equation}
    
    \noindent where $b_{max}$ and $b_{min}$ are upper and lower cutoffs on the impact parameter for Coulomb scattering~\cite{Lee1984, Ramis2012}. Eq.~\ref{lnLambda} ensures the avoidance of negative values for the Coulomb logarithm at low temperatures. The $b_{max}$ is set by the screening effect or Debye-Hückel~\cite{eliezer,Lee1984}. Considering the effect of electron degeneracy, the Debye-Hückel length of an electron distribution can be approximated by

    \begin{equation}
        \begin{split}
            \frac{1}{b_{max}^{2}} = \frac{1}{\lambda_{DH}^{2}} & = \frac{1}{\lambda^{2}_{De}} + \frac{1}{\lambda^{2}_{Di}},\\
            & = \frac{4\pi n_e e^2}{k_B(T_e^2 + T_F^2)^{1/2}} + \frac{4\pi n_i Z^2 e^2}{k_B T_i},
            \end{split}
        \label{LambdaD}
    \end{equation}

    \noindent where $\lambda_{Di}$ is the ionic Debye shielding length, and $n_i$ is the ion density~\cite{Lee1984,Brysk_1975}. In dense plasmas, the Eq.~\ref{LambdaD} breaks down as the screening length ($\lambda_{DH}$) becomes less than inter-atomic distances $R_0 = (3/4\pi n_i)^{1/3}$~\cite{Lee1984}. To avoid this nonphysical phenomenon, $R_0$ was set as the minimum cutoff for the electron mean free path $\lambda_e$~\cite{Lee1984, Eidmann00}. Accordingly, the cutoff collision frequency is given as:
    
     \begin{equation}
         \nu_{\mathrm{cutoff}} = \frac{\mathrm{electron \ total \ velocity}}{\mathrm{ion \ sphere \ radius}} = \frac{\sqrt{v_{th}^2+v_F^2}}{R_0}.
       \label{eq:cutoffFreq}
     \end{equation}
     
     \noindent The lower cutoff $b_{min}$ can be set by $\lambda_{dB}$ at high energy, according to the Heisenberg uncertainty principle, or by the classical distance of closest approach $Ze^2/3k_BT_e$ at low energy~\cite{Lee1984, Brysk_1975}. Taking into account the electron degeneracy and a partial ionization in high-density degenerate plasmas, $b_{min}$ can be reasonably approximated by~\cite{Brysk_1975}
    
    \begin{equation}
        b_{min} = \max \Bigg[ \frac{h}{2\sqrt{3m_ek_B \sqrt{T_e^2+T_F^2}}}, \frac{Ze^2}{3k_B\sqrt{T_e^2+T_F^2}} \Bigg].
        \label{bmin}
    \end{equation}
     
     The blue and red solid lines in Fig.~\ref{fig:collFreqPlots}~a) illustrate the temperature-dependent interpolated collision frequency $\nu_{interp}$ for electron-ion (ei) and electron-electron (ee) collisions implemented in 1D~PICLS code for the pre-ionized aluminium ($\mathrm{Al^{5+}}$). $T_F$ was considered as the threshold temperature for the degenerate plasma regime, and $\nu_{el-ph}$ was calculated assuming the local thermodynamic equilibrium (LTE) condition $T_e = T_i$. 
     
     \noindent The evolution of the electron and ion temperature of $\mathrm{Al^{5+}}$ plasmas using our new collision frequency model in PIC simulation is summarized in Fig.~\ref{fig:collFreqPlots}~b) with solid lines. The initial electron and ion temperatures were set to 58 and 0.5~eV, respectively, and the ionization was turned off. The system is confined between periodic boundaries, and no temperature gradient or heat diffusion is assumed. The thermalization dynamics are compared with the simple analytical calculation (dashed lines), where the relaxation rate is determined by the difference in electron and ion temperatures ~\cite{kodanova2018}:

     \begin{equation}
        \begin{split}
            & \frac{dT_e}{dt} = -\frac{n_i}{n_e} \frac{T_i-T_e}{\tau_{ei}},\\
            & \frac{dT_i}{dt} = \frac{T_e-T_i}{\tau_{ei}},
            \end{split}
        \label{eq:thermalization}
    \end{equation}

    \noindent where $\tau_{ei}$ is the characteristic time of ion heating~\cite{Eidmann00};
     
     \begin{equation}
         \tau_{ei} = \frac{m_i}{2m_e \nu_{ei}},
         \label{eq:tauheating}
     \end{equation}
     
     \noindent with $m_i$ denoting the ion mass. The electron-ion collision frequency was chosen as $\nu_{ei} = 1.8 \times 10^{16}$~$\mathrm{s^{-1}}$, to be consistent with its value in PIC simulation at 58~eV. Regarding the ion heating, a notable agreement between the PIC simulation and the analytical predictions is observed. However, the electron temperature exhibits a more rapid decay in the PIC simulation due, likely, to the radiation loss in the PIC simulation, a topic to be discussed in greater detail in the subsequent section.\\

     It is worth emphasizing that although the collision model correction is demonstrated in a one-dimensional PIC code, the momentum transfer accounting that results from collisions is performed in a comprehensive three-dimensional velocity space.
        
\section{SIMULATION RESULTS}
    We tested the constant $\nu_{const}$ and interpolated $\nu_{interp}$ collision frequency models in a 1D PIC simulation of the laser-excited Al thin foil. The simulation includes ionization dynamics and non-LTE conditions. This pure PIC simulation does not incorporate quantum effects or electron degeneracy within the algorithm of the PIC framework. The sample is a 200~nm thick aluminium ($\mathrm{Al^{3+}}$) foil at ambient temperature initially ($T_e=T_i=300$~K). A p-polarized laser with 50~fs duration at full-width half maximum (FWHM), 800~nm wavelength, and $6 \times 10^{14}\,\rm W/cm^2$ peak intensity irradiates on the Al sample surface.  
    To resolve $\lambda_e \sim 3$~\AA~around $T_F$, the cell size was set to be $\Delta x = 1.25$~\AA~corresponding to the time-step of $\Delta t = 4.16 \times 10^{-19}$~s. Each cell initially contains 12 ions ($\mathrm{Al^{3+}}$) and 36 free electrons. We also used the fourth-order particle shape to minimize the numerical noise~\cite{birdsall2018}. The ion number density was set to the density of Al ($n_{Al} = 34.6n_c$), where $n_c = m_e \omega_L^2/(4 \pi e^2) = 1.742 \times 10^{21} \mathrm{cm^{-3}}$ is the critical plasma density at the laser wavelength of $\lambda_L = 800$~nm. Here, $\omega_L$ is the laser angular frequency. The ionization dynamics due to the intense laser EM-field and the electron-ion collisions were modeled by the Landau-Lifshitz field and direct-impact (DI) ionization models, respectively.\\
    
    For the comparison, we also performed a similar simulation using the 1D~MULTI-fs hydrodynamic simulation code, specifically designed for short ($\leq$ ps) pulse high-intensity laser-solid interactions below $< 10^{17} \,\mathrm{W/cm^2}$ laser intensity~\cite{Eidmann00, *Ramis2012}. MULTI-fs calculates an explicit solution to Maxwell's equations for the fs laser pulse interactions with steep plasma density gradients. It includes the temperature-dependent collision frequency as well as a thermal conductivity from metallic solids up to high-temperature ideal plasmas and a separate equation-of-state (EOS) for electrons and ions (two-temperature model). We used EOS and ionization tables for Al provided by the MULTI-fs package. 500 cells were used for simulating the 200~nm thick Al foil. Finer cells were used closer to the front and rear surfaces compared to the middle of the foil to reduce the numerical error. The electron collision frequency was modeled using the same interpolated collision frequency model as implemented in the PICLS code. The radiation transport module was switched off, and the heat flux inhibition parameter was set to f = 0.1, which is defined as the ratio between the actual heating flux and the free streaming flux. The latter refers to the heat flow carried by electrons when they travel freely at their thermal velocity along the temperature gradient~\cite{eliezer}. The value of \textit{f} equal to 0.1 has been chosen based on comparisons between hydrodynamic code calculations and experiments on heat wave propagation into solids, which have shown good agreement for this value of \textit{f}~\cite{guethlein1996, Eidmann00}. \\
    
    \begin{figure}
         \centering
         \includegraphics[width=1\linewidth]{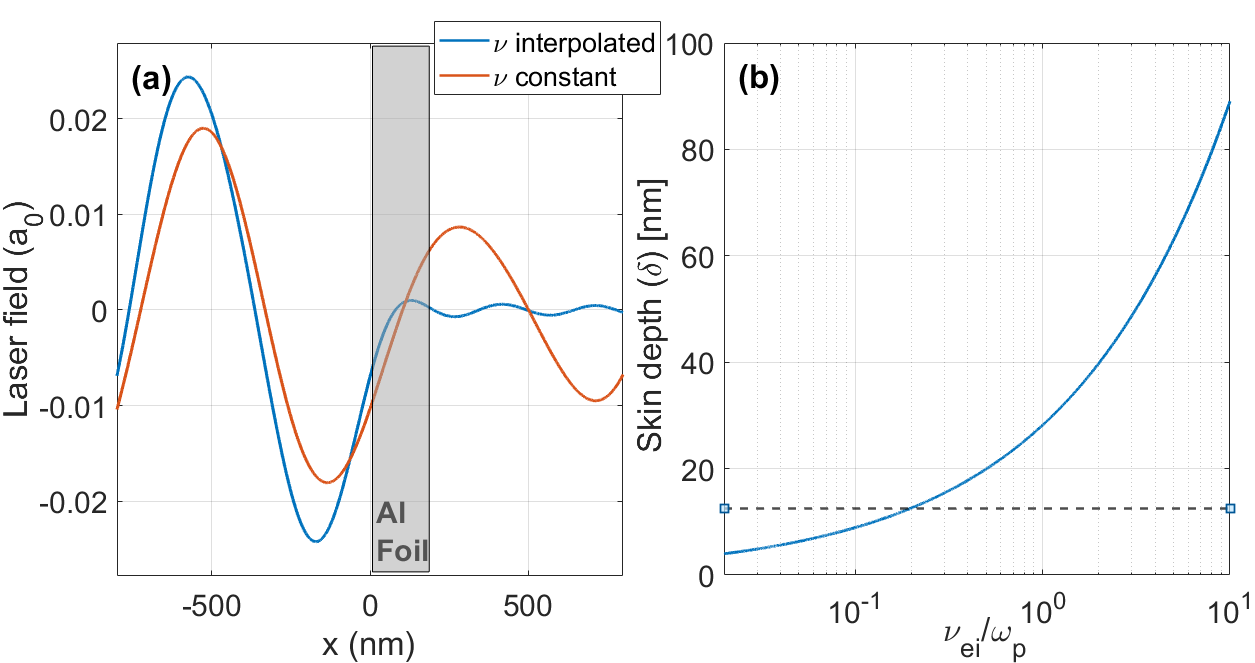}
         \vspace*{-6mm}
         \caption{(a) The amplitude of the transverse oscillating electric field vs. depth ($x$) at the laser temporal peak extracted from the PIC simulation employing the interpolated ($\nu_{const}$, blue) and constant ($\nu_{interp}$, red) collision frequency models. The target is a 200~nm Al foil with $Z=3$ initial ionization state. The front surface of the foil is located at $x=0$. A laser pulse with a 50~fs (FWHM) duration, 800~nm wavelength, p-polarization, and $6 \times 10^{14}\,\rm W/cm^2$ peak intensity impinges on the Al foil surface from the left side. The laser peak intensity reaches the front surface at $t=0$~ps. (b) Collisional skin depth vs. $\nu_{ei}$, for the propagation of an EM-wave with a wavelength of 800~nm inside a solid-density $\mathrm{Al}^{+3}$ plasma. The horizontal dashed line indicates the collisionless skin depth ($\delta_{collLess} = c/\omega_p$).}
         \label{fig:EMdesnityColl}
     \end{figure}
     
    Fig.~\ref{fig:EMdesnityColl}~a) illustrates the amplitude of the transverse oscillating electric field of the laser ($E_y$) as a function of the position in 1D PIC simulations with $\nu_{const}$ and $\nu_{interp}$ models when the peak laser intensity reaches the Al foil front surface on the left side. The grey shaded area indicates the foil. For the $\nu_{const}$ case, there is considerable leakage of the laser field due to the overestimation of $\nu$ at low temperatures. This is caused by the suppression of reflection due to excessive collisions  ($\lambda_e < 1$~nm), which prevent electrons from coherent oscillations in the laser field. This increases the collisional skin depth as
    
    \begin{equation}
         \delta = \frac{i}{\sqrt{\varepsilon}} \frac{c}{\omega_L},
         \label{eq:SkinDepth}
     \end{equation}
     
     \noindent where $\varepsilon$ is the dielectric function of the plasma medium defined by~\cite{eliezer}
     
     \begin{equation}
         \varepsilon = 1-\frac{\omega_p^2}{\omega_L^2(1+\mathrm{i}\nu_{ei}/\omega_L)}.
         \label{eq:pngilon}
     \end{equation}
     
    \noindent Fig.~\ref{fig:EMdesnityColl}~b) illustrates the skin depth as a function of $\nu_{ei}$, for the propagation of an EM-wave with a wavelength of 800~nm inside a solid-density $\mathrm{Al}^{+3}$ plasma. The skin depth increases with $\nu_{ei}$, resulting in a more extended penetration of the EM field into solids. The horizontal dashed line represents the collisionless skin depth ($\delta_{collLess} = c/\omega_p$). On the other hand, the laser EM-field is highly shielded by high-frequency plasma oscillations in the case with $\nu_{interp}$. As a result, the laser is reflected more from the surface and is highly damped inside solids.\\
    
    \begin{figure*}
         \centering
         \begin{subfigure}{}
           \includegraphics[width=1\linewidth]{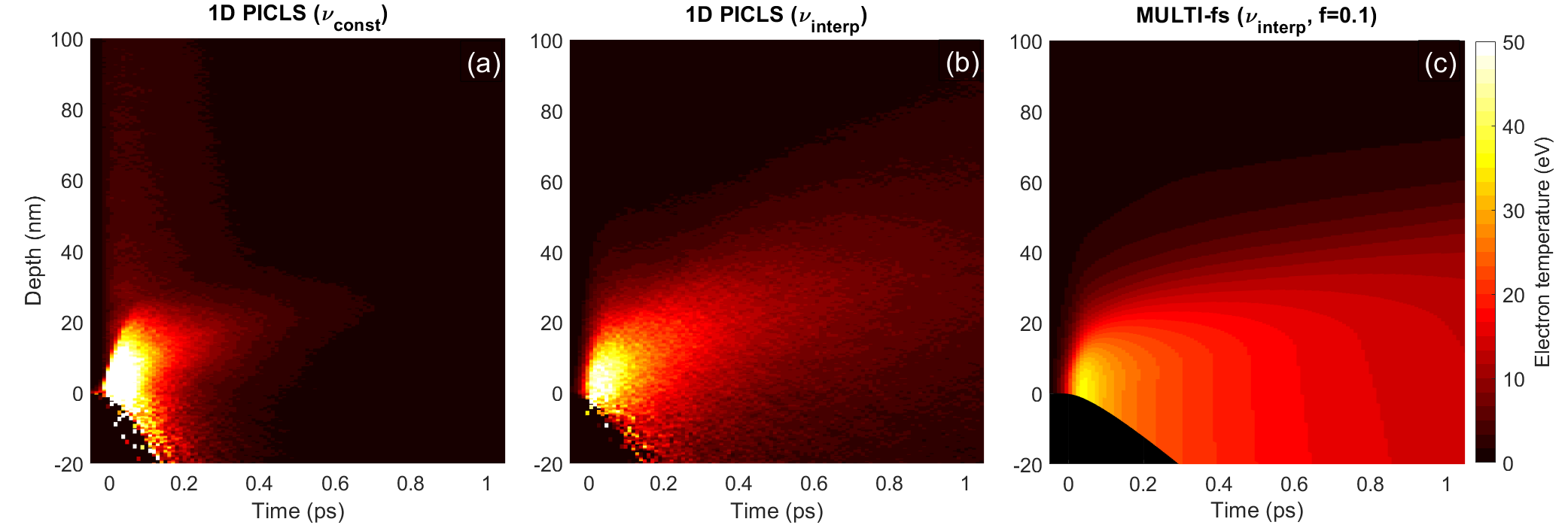}
        \end{subfigure}\\
        \vspace*{-10mm}
        \begin{subfigure}{}
           \includegraphics[width=1\linewidth]{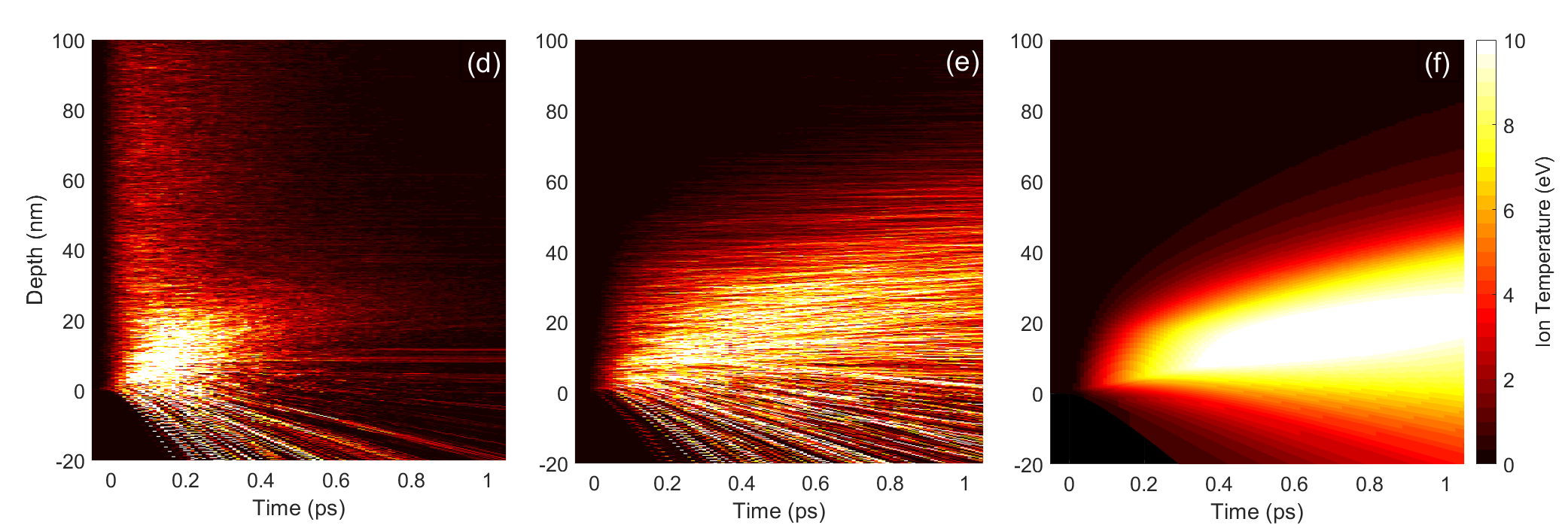}
        \end{subfigure}
         \vspace*{-6mm}
         \caption{Electron temperature ($T_e$) 2D maps as a function of time (horizontal axis) and depth (vertical axis), in the PIC simulation with $\nu_{const}$ (a) and $\nu_{interp}$ (b) collision models, and MULTI-fs (c). Target and the laser parameters in all simulations are the same as in fig.~\ref{fig:EMdesnityColl}. Figs.~(d-f) are the corresponding ion temperature 2D maps. Time and depth = 0 correspond to the peak intensity of the laser pulse and initial front surface of the target, respectively.  $T_i$ in PIC simulations were calculated using the local average kinetic energy of ions in the transverse direction. The radiation transport module was switched off and the heat flux inhibition parameter was set to f = 0.1in MULTI-fs.}
         \label{fig:temperatures}
     \end{figure*}
    
    Fig.~\ref{fig:temperatures} summarized the retrieved electron (a-c) and ion (d-f) temperature profiles inside the Al foil as a function of time and depth in 1D~PIC with $\nu_{const}$ (left), $\nu_{interp}$ (middle), as well as the MULTI-fs (right). The zero time and depth correspond to the peak intensity of the laser pulse and the initial front surface of the target, respectively. Due to the short duration of the laser pulse and the high density of the Al foil target, the surface expansion during laser irradiation is almost negligible. The laser's energy is deposited primarily at a solid density and instantly elevates the electron temperature at the surface, resulting in well-separated electron and ion temperatures. In this laser intensity regime, the main absorption mechanism in metallic solids is the inverse Bremsstrahlung, which directly depends on the collision frequency~\cite{eliezer}. The laser absorption observed in the PIC simulation with $\nu_{interp}$ is around 29\% and 35\% in MULTI-fs, which are close to experimental measurements reported in~\cite{Price1995}. To compare with the theory, the absorption coefficient of a solid surface $A$, with a reflection coefficient of $R$, irradiated by a linearly polarized monochromatic EM wave is given by~\cite{eliezer} 
    
    \begin{equation}
         A = 1 - R, \qquad R = \bigg|\frac{1-\sqrt{\varepsilon}}{1+\sqrt{\varepsilon}}\bigg|^2.
       \label{eq:absCoeff}
     \end{equation}
     
    \noindent For $\nu_{ei} \approx 1. \times 10^{16}$~$\mathrm{s^{-1}}$, the Eq.~\ref{eq:absCoeff} yields the absorption coefficients of $A \approx 0.3$ for the ionization state in PIC with $\nu_{interp}$ ($\mathrm{Al}^{6.5+}$) and $A \approx 0.35$ for ionization state in MULTI-fs ($\mathrm{Al}^{4.5+}$), which agree with values measured by simulations. In PIC with $\nu_{const}$, 28\% of the incident light was transmitted, and 35\% was reflected from the front surface. Thus, one can estimate a 37\% laser energy deposition into the target. The higher absorption for $\nu_{const}$ can be attributed to the broader skin depth area and absorption of the laser in deeper parts of the target. Both PIC simulations also show a faster surface expansion due to the higher temperature and, therefore, higher pressure ($P=P_e+P_i=k_B(n_eT_e+n_iT_i)$) inside the Al foil, compared to the MULTI-fs. 
    
    Comparing the electron temperature $T_e$ in the PIC simulation with $\nu_{interp}$ and MULTI-fs (b and c), one can see a generally good agreement in the surface heating and subsequent heat transport into the target. During laser excitation, the surface temperature reaches around 40~eV in MULTI-fs, and 50~eV in PIC with a spatial gradient inside the target of size $\sim 18$~nm in both simulations, which agrees with the collisional skin-depth $c/\omega_p\sqrt{\nu_{ei}/2\omega_L}$~\cite{eliezer}, for $\mathrm{Al^{3+}}$ and $\nu_{ei}=10^{16}$~$\mathrm{s^{-1}}$. The higher $T_e$ in the PIC simulation at around 0~ps can be attributed to kinetic effects (e.g., electron oscillations in the laser field), which are excluded in the MULTI-fs simulation. The temperature gradient is quickly smoothed out by heat transfer via thermal conduction. Bulk heating due to the ballistic transport of oscillating electrons does not play an essential role here due to high collisionality with sub-nm $\lambda_e$. Therefore, the dominant energy transport is driven by the heat diffusion expressed by the two-temperature energy conservation equations as
     
    \begin{equation}
         \begin{split}
         & C_e \frac{\partial T_e}{\partial t} = -\nabla \cdot \bm{q} - \gamma(T_e - T_i) + Q(z,t),\\
         & C_i \frac{\partial T_i}{\partial t} = \gamma(T_e - T_i),
         \end{split}
         \label{eq:heat_diffusion}
    \end{equation}
    
    \noindent where $C_e$, $C_i$ are, respectively, the volumetric heat capacity of electrons and ions, $\bm{q} = \kappa \nabla T_e$ describes the electron thermal heat flow with $\kappa(\nu)$ being the heat conductivity, and $Q(z,t) = \partial I_{abs}⁄\partial z$ is the power density deposited by the laser with $I_{abs}$ being the absorbed laser flux. The energy transfer rate from electrons to ions is expressed by $\gamma = C_i \tau_{ei}^{-1}$. \\
    
    \begin{figure}
          \centering
          \includegraphics[width=1\linewidth]{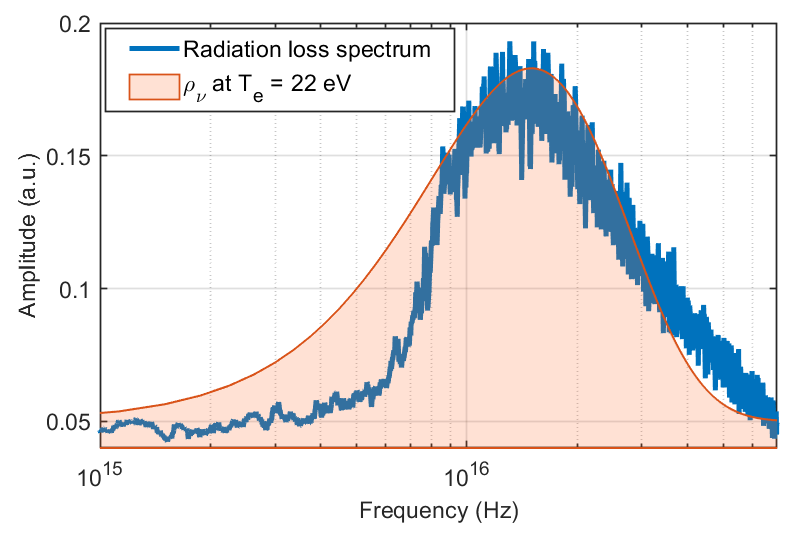}
          \caption{The spectrum of the radiative loss at the right boundaries (behind the target) extracted from the PIC simulation after laser excitation (blue line), and the Planckian electromagnetic energy density spectrum $\rho_\nu$ at $T_e = 22 eV$ (red-shade area).}
          \label{fig:spectrum}
    \end{figure}
    
    \noindent Both simulations also exhibit a similar depth of heatwave propagation. However, PIC with $\nu_{interp}$ shows a relatively faster electron cooling than MULTI-fs. In PIC, a considerable part of such fast electron cooling can be attributed to the collisional ionization process (roughly 28\% of the absorbed laser energy is used to ionize bound electrons). We also observed that almost one-third of the electron kinetic energy was left from the simulation box due to radiation energy loss via Bremsstrahlung. The blue line in fig.~\ref{fig:spectrum} illustrates the spectrum of radiation loss at the rear side of the target in PIC with $\nu_{interp}$. The spectrum peak intensity is around $1.5\times10^{16}$~Hz, which coincides with the Planckian electromagnetic energy density $\rho_\nu$ at $T_e = 22 eV$, as shown by the red area. This temperature is close to that observed in PIC, up to 1~ps. In addition, one can see a lower radiation power loss at frequencies below $10^{16}$~Hz compared to the Planckian distribution due to plasma shielding.
    
    \noindent In PIC simulation with $\nu_{const}$, the overestimated collision frequency and the excessive leakage of the laser field lead to a higher temperature (4-5~eV) in deeper parts of the target during laser excitation, compared to two other simulations. The overestimation of collisional heating also results in a higher surface temperature than in other cases. However, the exceeding electron collision frequency leads to very limited heat conduction into the target after the laser excitation~\cite{eliezer}. We also observed a higher radiation loss compared to the case with $\nu_{interp}$, resulting in faster surface cooling.
    It is useful to mention that the highest level of agreement with MULTI-fs simulations was achieved for the empirical parameter $K_s = 78.8$ for electron-phonon collisions in the PIC simulation. We observed much-limited heat conduction for lower $K_s$ values in PIC simulations.\\
    
    The evolution of the ion temperature $T_i$ via electron-ion energy transfer is summarized in Figs.~\ref{fig:temperatures}~(d-f). To focus on the heating via collisions, only the \textit{transverse} kinetic energy, \textit{i.e.}, perpendicular to laser propagation is shown. Here, one can also see that the overall profile of the ion temperatures agrees between MULTI-fs and 1D~PICLS with $\nu_{interp}$. The maximum ion temperature of around 10~eV is reached in both simulations. It is also visible that the PICLS shows a faster ion cooling, which is likely to be attributed to the overestimated radiative energy loss. On the other hand, PICLS with $\nu_{const}$ (Fig.~\ref{fig:temperatures}~d)) shows a faster ion heating \textit{and} faster cooling. The ion temperature quickly reaches above 10~eV near the surface and 4~eV in deeper parts already during laser excitation due to the overestimation of the $\nu_{ei}$. However, the temperature decays very quickly via excessive collisions and radiative loss.\\
    
\section{COMPARISON WITH THE EXPERIMENT AND DISCUSSION}  
    In this section, we discuss how the improved collision model in PIC simulation better describes the nanoscale solid-density plasma dynamics around the Fermi temperature by comparing it with state-of-the-art experimental results ~\cite{randolph20}.

    \begin{figure}
         \centering
         \begin{subfigure}{}
           \includegraphics[width=1\linewidth]{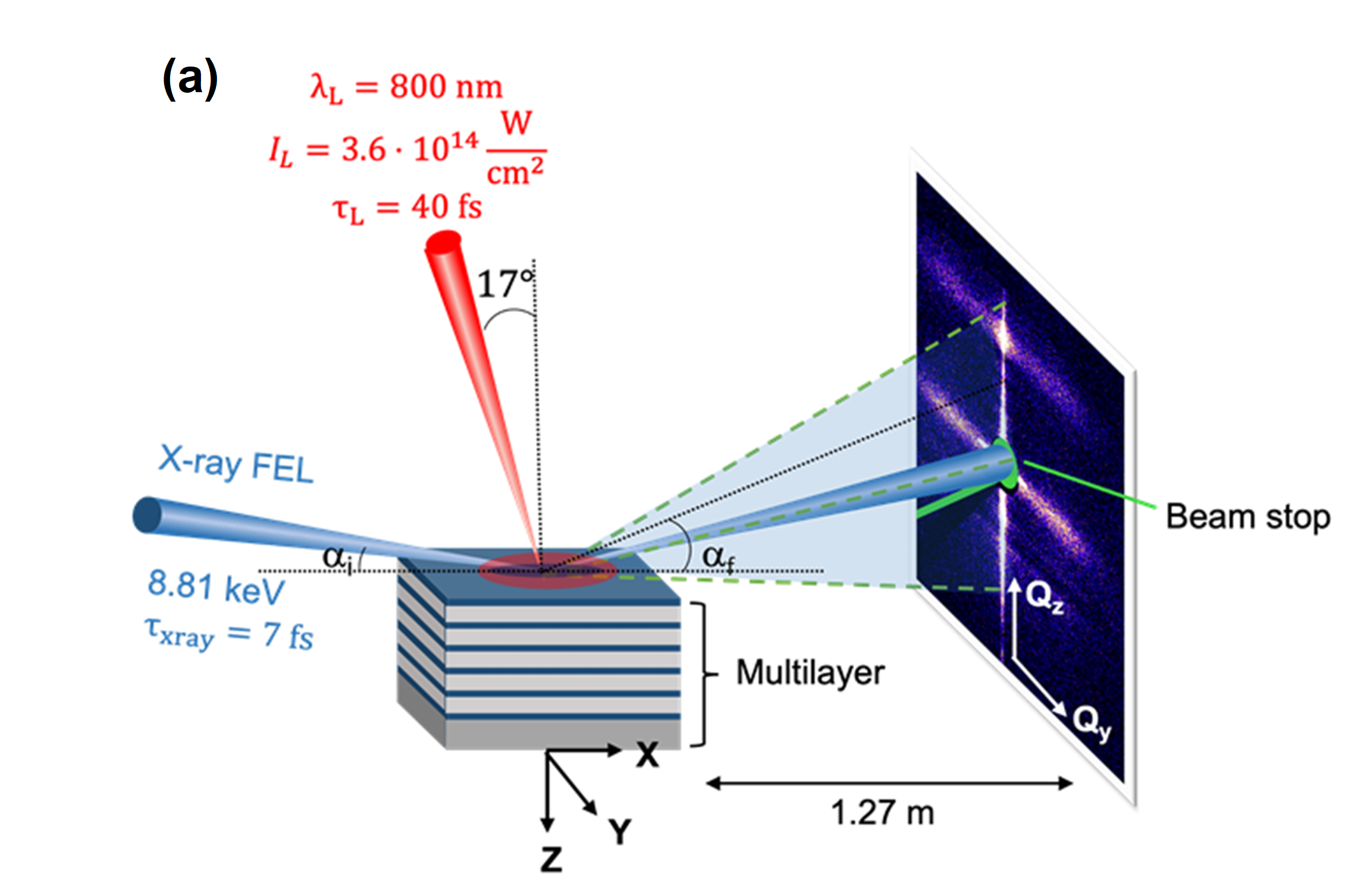}
        \end{subfigure}\\
        \vspace*{-5mm}
        \begin{subfigure}{}
           \includegraphics[width=1\linewidth]{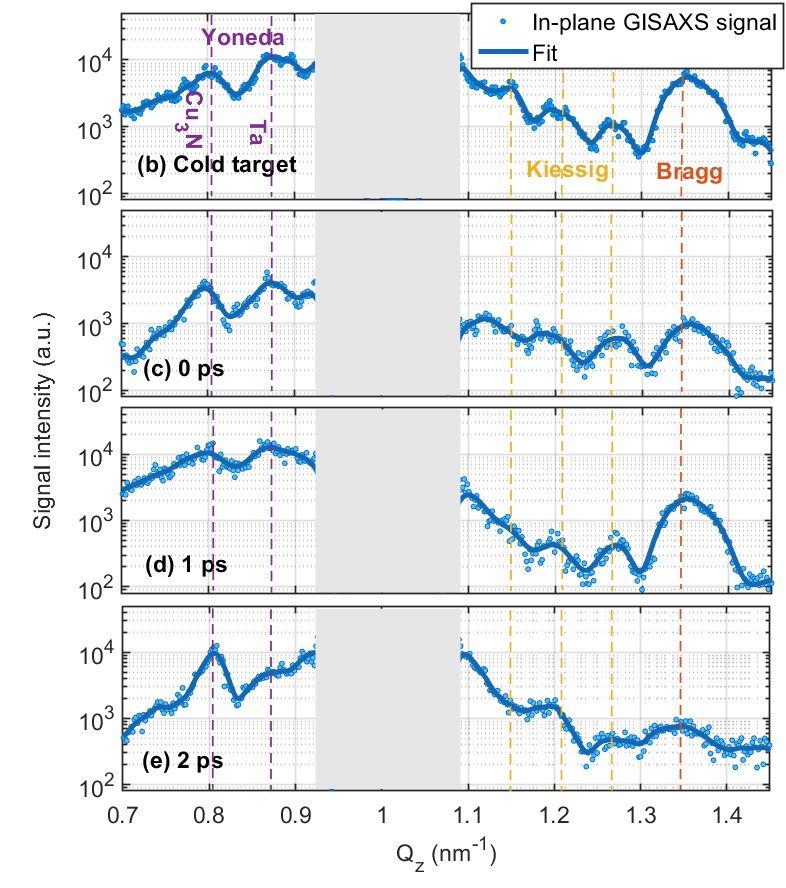}
        \end{subfigure}
         \vspace*{-6mm}
         \caption{(a) Schematics of the experimental setup at SACLA XFEL to investigate the subsurface solid-density plasma dynamics with grazing-incidence small-angle x-ray scattering (GISAXS)~\cite{randolph20, LisaThesis}. (b) Lineout of the in-plane scattering signal from the cold Ta-$\mathrm{Cu3N}$ multilayer target. (c-e) Lineout of the in-plane scattering signal after laser excitation at different delays. Scattered dots are the actual experimental signal, and solid lines represent fits. The vertical red, orange, and purple lines are the initial positions of the Bragg peaks, Kiessig fringes, and Yoneda peaks, respectively.}
         \label{fig:SACLA_Setup}
     \end{figure}
    
    The experiment was performed at the EH6 station at the SACLA X-ray free-electron laser (XFEL) facility in Japan~\cite{Yabuuchi19}. 
    A schematic of the experimental setup is shown in Fig.~\ref{fig:SACLA_Setup}~a). 
    A p-polarized Ti:sapphire short-pulse laser with 800~nm central wavelength, 70~mJ pulse energy, 40~fs FWHM duration was irradiating with  $17^{\circ}$ incident angle from the surface normal on multilayer (ML) samples. The ML samples consisted of 5 layers of tantalum ($\mathrm{Ta}$) and copper nitride ($\mathrm{Cu_3N}$) 4.3 / 11.5~nm thickness each. After a variable delay, an x-ray probe pulse with 7~fs duration and photon energy of 8.81~keV was focused to a 4~$\mu$m FWHM on the sample at a grazing incidence angle of $0.64^{\circ}$, which is slightly larger than the critical angle of total external reflection. Scattered photons were collected by a 2D area detector, multi-port charge-coupled device (MPCCD)~\cite{Kameshima14}. The strong specular peak at the exit angle equal to the incidence angle was blocked. Due to the grazing incidence geometry, the x-ray footprint on the surface was elongated in one dimension to $\sim 360 \mathrm{\mu m}$. Accordingly, the laser beam was defocussed to $\sim$ 500~$\mathrm{\mu m}$ diameter to cover the x-ray footprint, resulting in a laser intensity of $3.6\pm0.2\times10^{14}$~$\mathrm{W/cm^2}$. The x-ray scattering, as a consequence of the large probe area, represents a time-averaged signal of about 1.2~ps \cite{randolph20}. \\
        
    Fig.~\ref{fig:SACLA_Setup}~b-e) summarized the lineout of the x-ray scattering intensity along the plane spanned by the incoming x-ray beam and the target surface normal, denoted as $Q_z$ in Fig~\ref{fig:SACLA_Setup}~a), at different delays with respect to the laser intensity peak. 
    The circular dots represent the experimental data, while the solid lines are fits. This so-called in-plane scattering signal contains momentum transfers dominantly along the depth direction (z), which is closely associated with the specular reflectivity curve~\cite{Holy93}. The Bragg-like peak at $Q_z = 1.33 \,\mathrm{nm}^{-1}$ represents the typical length scale, \textit{i.e.} 15.8~nm thickness of each Ta/$\mathrm{Cu_3 N}$ double-layer. The small peaks between $Q_z = 1.0 \,\mathrm{nm}^{-1}$ and $1.33 \,\mathrm{nm}^{-1}$, are known as Kiessig fringes~\cite{Kiessig31}, resembling the number of double-layer repeats in the ML sample. The scattering signal at the exit angle below the incident angle $Q_z < Q_{\text{specular}}$ is particularly surface sensitive, where the evanescent x-ray wave travels parallel to the surface. The peaks at $0.80 \,\mathrm{nm}^{-1}$ and $0.87 \,\mathrm{nm}^{-1}$ (dashed vertical purple lines in Fig.~\ref{fig:SACLA_Setup}~b-e)) correspond to the solid densities of $\mathrm{Cu_3 N}$ and Ta, respectively, so-called Yoneda peaks~\cite{Yoneda63}. Those signals originate primarily from the interference of the topmost surface layer and serve as a sensitive marker for the structure of the uppermost layers. 

    \begin{figure*}
         \centering
         \includegraphics[width=1\linewidth]{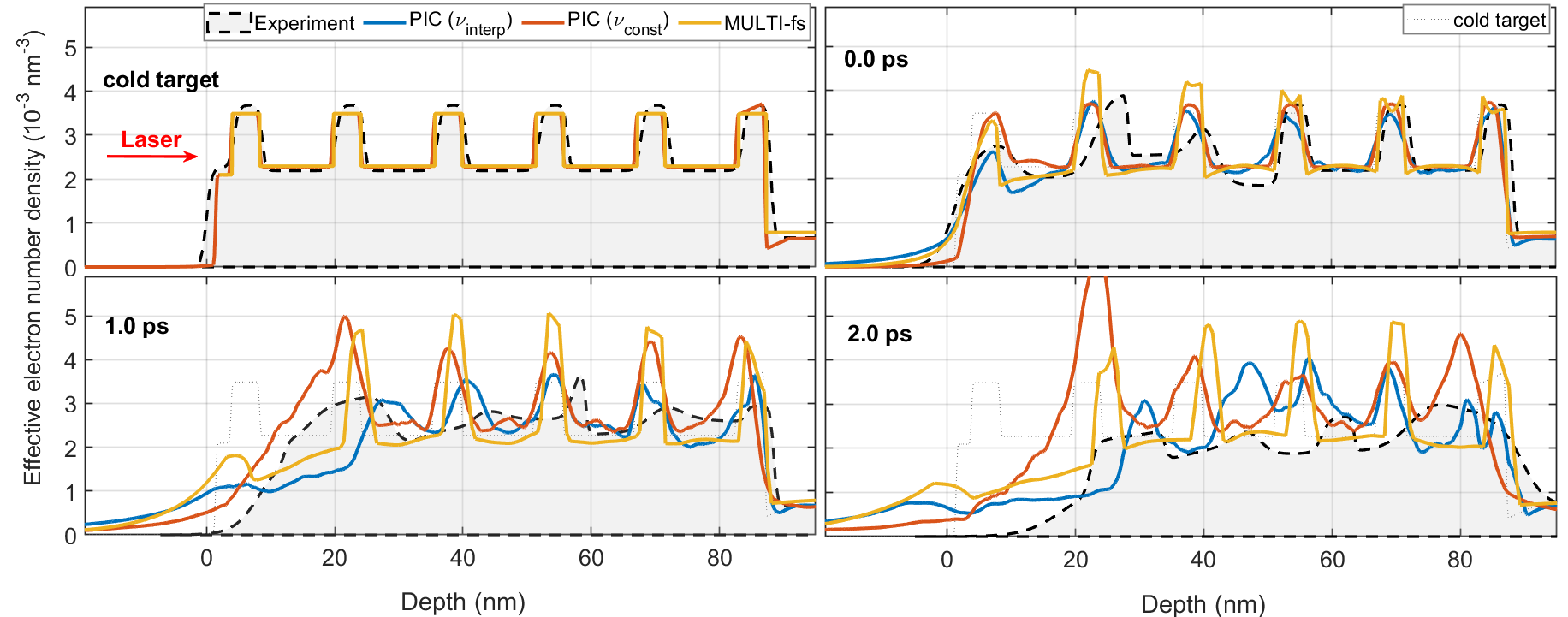}
         \vspace*{-7mm}
         \caption{Evolution of the effective electron density $n_{eff}$ retrieved from the experimental GISAXS data (the grey area with the dashed black lines), PIC simulation with $\nu_{interp}$ (blue) and $\nu_{const}$ (red) collision frequency models at low temperatures, and MULTI-fs simulation (yellow), at different delay times after laser irradiation. In simulations, the target was a Ta-Cu multilayer (ML) with a similar structure and size as the experiment. The Si substrate of the experimental target was replaced by a 200~nm-thick Al foil in simulations. The target was irradiated by a 50~fs (FWHM) laser pulse with 800~nm wavelength and $4 \times10^{14}$~$\mathrm{W/cm^2}$ peak intensity. The laser impinges perpendicularly on the ML surface from the left, and time = zero corresponds to the peak intensity of the laser pulse. The simulation results are averaged over 1.2 ps to be consistent with the experimental temporal resolution conditions. The dotted grey line shows the initial ML structure and position in simulations.}
         \label{fig:ExpandSimDesinty}
     \end{figure*}
    
    The change in the scattering signal is already visible at 0~ps (which is the integration over $\pm 0.60$~ps around the laser intensity peak) compared to that of the cold sample in the Yoneda peaks region, but also in the number of Kiessig fringes and the Bragg peak intensity. Those modulations are more pronounced in later times; 1~and~2~ps. 
    By comparing these x-ray scattering signals with the reconstructed electron density profile, shown as grey-shaded areas with the dashed black lines in Fig.~\ref{fig:ExpandSimDesinty}, this can be understood as the surface expansion (ablation) of the top Ta layer into vacuum and modulation of inner layers which intermix to each other. The details on the reconstruction from the x-ray scattering pattern to the real-space density profiles are discussed in detail in ref.~\cite{randolph20}. Here, we focus on the comparison between the experimental data and simulations.  
    Note that in our experiment, x-ray scattering was originated mostly from the bound electrons due to the high x-ray photon energy (8.81 keV) and relatively low mean ionization $Z_{mean} = 7~$\cite{Chung05}. The x-ray scatters from all electrons with binding energies below the photon energy, which corresponds to 27 electrons for Cu and 63 for Ta. This means we track here mostly the bound electrons or, equivalently, ion motion. We applied the same criteria for plotting the electron density (\textit{effective} electron density: $n_{eff}$) for our simulations. 
      
    Fig.~\ref{fig:ExpandSimDesinty} shows the comparison between the reconstructed density profiles from the experiment (the grey area with dashed black lines), PICLS using $\nu_{const}$ (red line) and $\nu_{interp}$ (blue line) and MULTI-fs (orange line). Each simulation result is averaged over 1.2~ps to be consistent with the time resolution of the experiment. 
    In Multi-fs, the simulation box included 1208 cells, with 833 cells for the ML and 375 cells for the Al substrate. Finer cells were used close to the layer interfaces to avoid the numerical noise. The EOS and ionization data for Ta and Cu were produced using FEOS code~\cite{Faik18}. The radiation transfer module was switched off in the simulation. The free streaming limiting factor $f$ of 0.6 was used here instead of $f=0.1$ used above, to be consistent with our previous publication~\cite{randolph20}. We also verified that the selection of $f$ between 0.1 and 0.6 led negligible impact on overall density dynamics. The incident angle of the laser was set to $17\degree$ from the target normal to be consistent with the experiment. For the substrate material, instead of silicon (Si) used in the experiment, aluminium (Al) was used in our simulations because of its well-known EOS. As the energy transport occurs from the ML to the substrate, no meaningful difference in ML dynamics between these two material choices is expected. Only the expansion of the last Ta layer into the substrate can be slightly different between these two materials due to the difference in their densities by about 20\%. \\
    In PICLS, each cell contains 15 virtual ion particles with initial charge states of 2, 1, and 3 for Ta, Cu, and Al, respectively. The fourth-order particle shape with different particle weightings for different ion species is used. The ion number densities are set to realistic densities of Ta ($n_{\mathrm{Ta}}=31.8n_c$), Cu ($n_{\mathrm{Cu}}=48.7n_c$), and Al ($n_{\mathrm{Al}}=34.6n_c$). The ionization dynamics are modeled using the field and direct-impact ionization models. Here, the incident angle of the laser was set to normal to the surface, as the oblique incidence is not supported in 1D PICLS. We verified from Multi-fs that the difference in laser absorption between normal-incidence and $17\degree$ incidence was less than 1\%, as has been reported previously~\cite{Eidmann00}. 
    For both simulations, the laser with $\tau=50$~fs duration, $\lambda_L=800$~nm wavelength, and $4\times10^{14} \mathrm{W/cm^2}$ intensity impinges on the surface of the ML target from the left side. \\
    It is important to note that the advantage of using the PIC here compared to the hydrodynamic simulation is, that despite its higher computational cost, the PIC allows simulating the particle exchanges between adjacent nm-scale layers, which leads to layer intermixing. This effect is excluded in Lagrangian hydrodynamic simulations as the boundary between cells needs to be preserved~\cite{Ramis2012, Eidmann00}. 

    \begin{figure}
         \centering
         \includegraphics[width=1\linewidth]{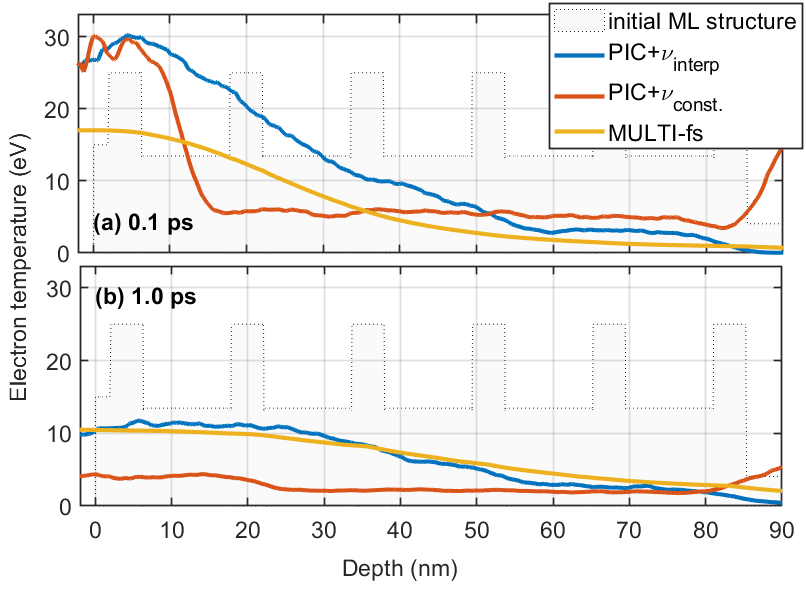}
         \vspace*{-7mm}
         \caption{The electron temperature profile inside the Ta-Cu ML target in the PIC simulation with $\nu_{interp}$ (blue), PIC with $\nu_{interp}$ (red), and MULTI-fs (orange). The top and bottom figures show the simulation results at 0.1 and 1~ps after the laser excitation, respectively. The grey area indicates the initial ML structure.}
         \label{fig:ML_Te}
     \end{figure}

    The electron temperature ($\mathrm{T_e}$) profiles in simulations at 0.1 and 1~ps after the laser-excitation are shown in figure~\ref{fig:ML_Te}a) and figure~\ref{fig:ML_Te}b), respectively. Here, different colors represent different simulations (same as fig.~\ref{fig:ExpandSimDesinty}). PIC simulations show a higher surface temperature than the MULTI-fs due to higher laser absorption, as already discussed in the previous section. The laser absorption in PIC with $\nu_{interp}$ was 43\%, which is in agreement with experimental measurements for Ta, reported in ref.~\cite{Price1995}, and was 29\% in the MULTI-fs. In PIC with $\nu_{const}$, 37\% of the laser energy was reflected from the surface, and around 13\% was transmitted through the target, including the ML and 200~nm Al substrate. This means around 50\% of the laser energy was deposited in the target. The overestimated laser propagation length into the target leads to a higher $\mathrm{T_e}$ in deeper parts (between 50 and 80 nm) of the ML target and the Al substrate. Furthermore, both PIC simulations here also show faster electron cooling than that observed in MULTI-fs due to excessive radiation energy loss.\\
    
    The highly localized surface heating instantaneously increases the pressure in the surface region. As a result, the top Ta layer expands rapidly into the vacuum within the first 1~ps, as can be seen in the experiment, PIC, and MULTI-fs simulations. The density modulation of the ML target starts from the surface and moves towards the substrate due to heat diffusion. The whole ML structure appears to be significantly modulated after $\sim$1~ps as indicated both by the experiment and simulations. 
    If we consider the position of $1/e$ of the surface temperature to be the position of the heat wavefront, the heatwave velocity of around 4-5$\times10^{4}$~$\mathrm{m/s}$ is observed in both MULTI-fs and PICLS with $\nu_{interp}$. This value was in good agreement with the ferocity of the density perturbation front observed in the experiment.
    
    \noindent In terms of the overall density dynamics, the PIC simulation with $\nu_{interp}$ shows a better agreement with the experiment compared to the PIC simulation with $\nu_{const}$. In MULTI-fs, the initial ML structure persists due to the lack of intermixing between Lagrangian cells. 
    
    The ablation of the solid-density surface develops over time as the pressure wave travels into the target. If we define the ablation surface as the position of half of the initial solid Cu density, the PIC simulation with $\nu_{interp}$ predicts an ablation speed of $\sim$10~nm/ps up to 2~ps. This is in good agreement with the experiment and also with the sound speed at $T_e=20$ and $T_i=5$~eV, assuming an ideal gas EOS~\cite{Drake} 
    
    \begin{equation}
        C_s = \Bigg( \frac{\gamma_e Z_{mean} k_B T_e + \gamma_i k_B T_i}{m_i} \Bigg)^{1/2}
        \label{soundSpeed}
    \end{equation}
    
    \noindent where $Z_{mean}$ is the mean ionization, $\gamma_e$ and $\gamma_i$ are adiabatic index of electrons and ions, respectively. $M_i=181$ a.u. and 63 a.u. is the ion mass for Ta and Cu, respectively, and $\gamma = 1+2⁄n$ with $n$ being the number of degrees of freedom. 
    In most cases, $\gamma_e = 1$ and $\gamma_i = 3$ can be used~\cite{Drake}. For ($T_e$,$T_i$)=(20,5)~ eV and $Z_{mean} = 5$, Eq.~\ref{soundSpeed} yields $C_{s \_ Ta} \sim 8 \, \mathrm{nm⁄ps}$ and $C_{s \_ Cu} \sim 13 \, \mathrm{nm⁄ps}$, respectively. 
    
    However, the PIC simulation with $\nu_{const}$ shows less ablation due to the faster surface cooling. 
     
    \begin{figure}
         \centering
         \includegraphics[width=1\linewidth]{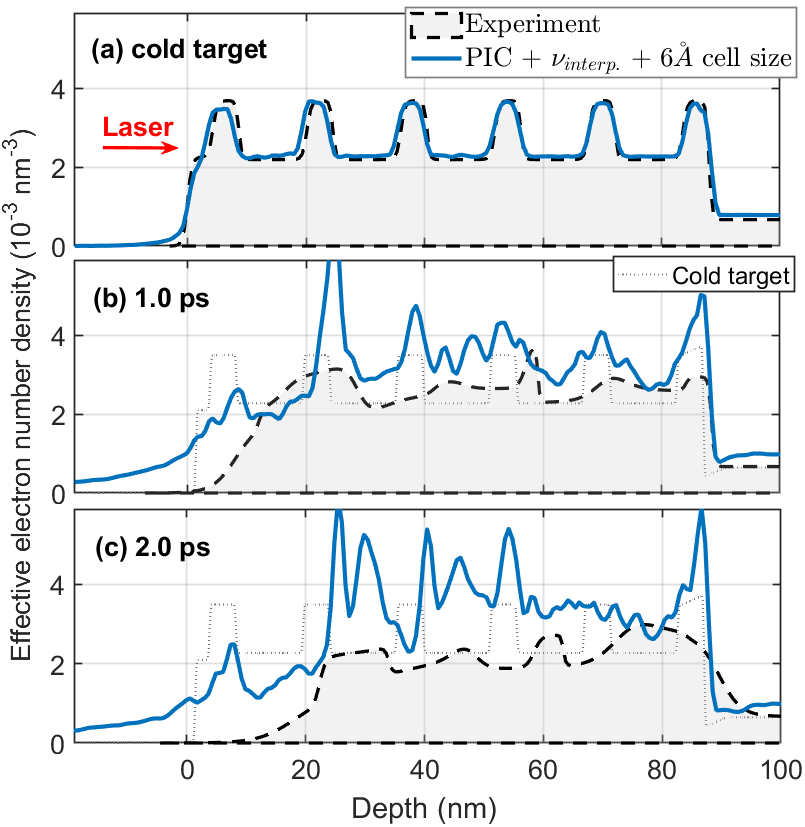}
         \vspace*{-5mm}
         \caption{Effective electron density as a function of depth in GISAXS experiment (grey area with dashed black line) and PIC simulations with $\nu_{interp}$ and larger grid size of $\Delta x = 6$~\AA~(solid blue line) at different delay times after laser irradiation. The laser and ML target parameters are the same as in Fig~\ref{fig:ExpandSimDesinty}.}
         \label{fig:lowResolution}
    \end{figure}
     
    Furthermore, we investigated the impact of the selection of the grid size on PIC simulation under the strongly coupled plasma regime.
    Fig.~\ref{fig:lowResolution} shows the result obtained with the larger grid size of 6~\AA, which is larger than the electron mean free path at $T_e=10$~eV; as the total electron velocity of $v_e=\sqrt{v_{th}^2+v_{F}^2}=2.7\times10^6$~$\mathrm{m/s}$ and the collision frequency $\nu \approx 1. \times 10^{16}$~$\mathrm{s^{-1}}$ leads $\lambda_e = 2.7$~\AA. This resulted in lower collisional absorption of 32\% compared to the case of 1.25~\AA~ grid size with 43\% absorption. This led to lower $T_e$, thus slower subsequent density modulation. Consequently, the agreement with the experiment is lost, as shown in~\ref{fig:lowResolution}~b-c). We found the maximum cell sizes should not exceed 1.5 times the electron mean free path. Also, the higher number of ions per cell results in more accurate and stable outputs.
     
\section{CONCLUSION ANd PERSPECTIVES}
    We demonstrated that employing a correct collision frequency around the Fermi temperature with the Angstrom-scale cell size significantly improves the predictive capability of the current PIC simulations at the degenerated plasma regime. It is revealed that our new model reasonably describes the density dynamics of the laser-excited nanometric ML samples at a few eV temperatures up to $\sim 2$ ps. However, the implemented collision model still suffers from several fundamental issues that may affect its predictive power. For instance, the binary collision model implemented in the PICLS code is suitable only for small-angle scattering, a reasonable assumption in high-temperature plasmas~\cite{SENTOKU2008} but not at low temperatures and high densities. This limitation becomes particularly significant when dealing with low-temperature dense plasmas, where the assumptions of the binary collision model may not hold. Here, one needs careful attention when interpreting results and making conclusions about the electron-ion equilibrium rate and other essential plasma properties.

    Moreover, the relatively substantial radiation loss, approximately $\sim30$ \% of the total energy, observed in the PIC simulation can result in a pronounced effect on the electron temperature and the subsequent density evolution. This is primarily attributed to Bremsstrahlung radiation — a consequence of electron collisional processes. To address this issue, incorporating a radiation transport module into the simulation is necessary. Such an enhancement would enable a more accurate characterization of the interplay between radiation loss and its impact on the dynamics of the laser-excited ML target.

    Our experimental results also indicate that the intermixing of the nanolayer within the ML target plays a crucial role in the density dynamics. This intermixing effect, however, is not fully captured in the current 1D PIC simulation employed in this paper. The 1D nature of the simulation may not be well-suited to capture the three-dimensional complexities of the system and its consequential impact on the density dynamics of laser-excited multilayer targets.
    
    Quantum mechanical effects and electron degeneracy, crucial factors in accurately describing the behavior of particles in low-temperature and high-density plasmas, are not inherently taken into account in PIC simulations. One may also encounter methodological challenges in exploring the parameter space studied here, which intersects with plasma parameters typically investigated using Time-Dependent Density Functional Theory (TDDFT)~\cite{dornheim2023} and Molecular Dynamics (MD)~\cite{bussmann2006, bussmann2008} simulations. These challenges are especially pronounced under conditions featuring exchange-correlation effects in cold and dense environments. The adoption of PIC simulations in this realm is innovative, and the contributions outlined here allow for the study of this parameter space using a simplified surrogate model approach. This approach builds upon state-of-the-art collisional operators and ionization models. However, it introduces complexity when diverging from an LTE framework towards a non-LTE approach. Specifically, it implicitly assumes LTE by constructing dynamical means from the local phase space distribution, a process that overlooks the long- and short-range order within the material. At the same time, it provides access to study large-scale systems inaccessible by TDDFT/MD methods. While enabling simulations at this scale, it may neglect physical interactions that could impact the accuracy of our findings, which can best be evaluated in comparison to experiments.

    Including the above-mentioned physics in our PIC simulation would require prohibitive computational cost, especially for a multidimensional PIC simulation with angstrom scale cell size. The numerical effort for solving Maxwell's equations scales as ($\propto 1/\Delta x^4$), making such an approach impractical~\cite{peltz2012}. This computational limitation can be effectively addressed by leveraging modern, stable, and accelerated PIC simulation frameworks like PIConGPU \cite{hubl2019, PIConGPU}. Additionally, innovative hybrid approaches, such as MicPIC~\cite{bart2017, peltz2012}, combining MD and PIC techniques, offer a promising strategy to overcome this challenge. In MicPIC, MD handles short-range electrostatic interactions, while PIC is employed for long-range electromagnetic effects, providing a balanced and efficient solution.

    Experimental measurements using GISAXS could serve as reliable benchmarks for assessing the impact of implementing new physics in PIC simulations. Here, some advancements in measurement techniques can be explored to enhance the utility of these experimental benchmarks and improve their synergy with simulations. Specifically, adopting nanofocusing of XFEL beams in experiments holds the potential to enhance temporal resolution significantly. This improvement enables more accurate and precise capture of ultra-fast density dynamics in laser-excited targets. When incorporated into simulations, these improved experimental results can offer a more detailed understanding of crucial processes, such as laser absorption and subsequent heat propagation into the target. The intricacies of these processes, inherently dependent on collisions, become more discernible with better experimental precision.
    
    Moreover, the improved experimental time-resolution precision provides a unique opportunity to investigate the mechanisms that initiate the intermixing process between adjacent nanolayers of dense plasmas in the early stages after the laser excitation. It also enables a closer examination of how collisional effects influence the speed and dynamics of the intermixing process. Consequently, the refined experimental benchmarks not only validate simulation outcomes but also facilitate a more nuanced exploration of the underlying physics, contributing to a deeper understanding of laser-solid interactions and subsequent plasma dynamics.

\section{ACKNOWLEDGMENTS}
    M.B., C.G., and M.Na. acknowledge funding by DFG GU 535/6-1. We are particularly grateful for the invaluable contributions of our colleagues. Special thanks to Y. Sentoku (Osaka University) for offering essential insights into PIC simulations. We extend our gratitude to R. E. Brambrink (EuXFEL) for his expert advice on hydrodynamic simulations, and we appreciate the valuable discussions with J.-P. Schwinkendorf (HZDR), L. Wollenweber (EuXFEL), M. Habibi (EuXFEL), C. F. Grote (Max Planck Institut für Evolutionsbiologie), and C. Rödel (U. Schmalkalden).
    The success of the XFEL experiments at BL2 of SACLA, approved by the Japan Synchrotron Radiation Research Institute (JASRI) under Proposal No. 2018B8049, was a collaborative effort directed by C.G. and M.Na. The team, including L.R., M.Ba., T.R.P., M.M., N.P.D., S.G., T.M., M.Ni., C.G., and M.Na., was ably supported by T. Yabuuchi, Y. Inubushi, K. Sueda, and T. Togashi. We also acknowledge N.P.D. and J.K.K. for their assistance in generating EOS tables. The manuscript benefited from the comments and suggestions of all involved.  

\bibliography{Ref_GISAXS2}

\begin{thebibliography}{54}%
\makeatletter
\providecommand \@ifxundefined [1]{%
 \@ifx{#1\undefined}
}%
\providecommand \@ifnum [1]{%
 \ifnum #1\expandafter \@firstoftwo
 \else \expandafter \@secondoftwo
 \fi
}%
\providecommand \@ifx [1]{%
 \ifx #1\expandafter \@firstoftwo
 \else \expandafter \@secondoftwo
 \fi
}%
\providecommand \natexlab [1]{#1}%
\providecommand \enquote  [1]{``#1''}%
\providecommand \bibnamefont  [1]{#1}%
\providecommand \bibfnamefont [1]{#1}%
\providecommand \citenamefont [1]{#1}%
\providecommand \href@noop [0]{\@secondoftwo}%
\providecommand \href [0]{\begingroup \@sanitize@url \@href}%
\providecommand \@href[1]{\@@startlink{#1}\@@href}%
\providecommand \@@href[1]{\endgroup#1\@@endlink}%
\providecommand \@sanitize@url [0]{\catcode `\\12\catcode `\$12\catcode `\&12\catcode `\#12\catcode `\^12\catcode `\_12\catcode `\%12\relax}%
\providecommand \@@startlink[1]{}%
\providecommand \@@endlink[0]{}%
\providecommand \url  [0]{\begingroup\@sanitize@url \@url }%
\providecommand \@url [1]{\endgroup\@href {#1}{\urlprefix }}%
\providecommand \urlprefix  [0]{URL }%
\providecommand \Eprint [0]{\href }%
\providecommand \doibase [0]{http://dx.doi.org/}%
\providecommand \selectlanguage [0]{\@gobble}%
\providecommand \bibinfo  [0]{\@secondoftwo}%
\providecommand \bibfield  [0]{\@secondoftwo}%
\providecommand \translation [1]{[#1]}%
\providecommand \BibitemOpen [0]{}%
\providecommand \bibitemStop [0]{}%
\providecommand \bibitemNoStop [0]{.\EOS\space}%
\providecommand \EOS [0]{\spacefactor3000\relax}%
\providecommand \BibitemShut  [1]{\csname bibitem#1\endcsname}%
\let\auto@bib@innerbib\@empty
%</preamble>
\bibitem [{\citenamefont {Lee}\ \emph {et~al.}(2002)\citenamefont {Lee}, \citenamefont {Baldis}, \citenamefont {Cauble}, \citenamefont {Landen}, \citenamefont {Wark}, \citenamefont {NG}, \citenamefont {Rose}, \citenamefont {Lewis}, \citenamefont {Riley}, \citenamefont {Gauthier},\ and\ \citenamefont {et~al.}}]{lee_baldis2002}%
  \BibitemOpen
  \bibfield  {author} {\bibinfo {author} {\bibfnamefont {R.W.}\ \bibnamefont {Lee}}, \bibinfo {author} {\bibfnamefont {H.A.}\ \bibnamefont {Baldis}}, \bibinfo {author} {\bibfnamefont {R.C.}\ \bibnamefont {Cauble}}, \bibinfo {author} {\bibfnamefont {O.L.}\ \bibnamefont {Landen}}, \bibinfo {author} {\bibfnamefont {J.S.}\ \bibnamefont {Wark}}, \bibinfo {author} {\bibfnamefont {A.}~\bibnamefont {NG}}, \bibinfo {author} {\bibfnamefont {S.J.}\ \bibnamefont {Rose}}, \bibinfo {author} {\bibfnamefont {C.}~\bibnamefont {Lewis}}, \bibinfo {author} {\bibfnamefont {D.}~\bibnamefont {Riley}}, \bibinfo {author} {\bibfnamefont {J.-C.}\ \bibnamefont {Gauthier}}, \ and\ \bibinfo {author} {\bibnamefont {et~al.}},\ }\bibfield  {title} {\enquote {\bibinfo {title} {Plasma-based studies with intense x-ray and particle beam sources},}\ }\href {\doibase 10.1017/S0263034602202293} {\bibfield  {journal} {\bibinfo  {journal} {Laser and Particle Beams}\ }\textbf {\bibinfo {volume} {20}},\ \bibinfo {pages} {527–536} (\bibinfo {year}
  {2002})}\BibitemShut {NoStop}%
\bibitem [{\citenamefont {Rightley}\ and\ \citenamefont {Baalrud}(2021{\natexlab{a}})}]{rightley2021}%
  \BibitemOpen
  \bibfield  {author} {\bibinfo {author} {\bibfnamefont {Shane}\ \bibnamefont {Rightley}}\ and\ \bibinfo {author} {\bibfnamefont {Scott~D}\ \bibnamefont {Baalrud}},\ }\bibfield  {title} {\enquote {\bibinfo {title} {Kinetic model for electron-ion transport in warm dense matter},}\ }\href@noop {} {\bibfield  {journal} {\bibinfo  {journal} {Physical Review E}\ }\textbf {\bibinfo {volume} {103}},\ \bibinfo {pages} {063206} (\bibinfo {year} {2021}{\natexlab{a}})}\BibitemShut {NoStop}%
\bibitem [{\citenamefont {Moldabekov}\ \emph {et~al.}(2015)\citenamefont {Moldabekov}, \citenamefont {Ludwig}, \citenamefont {Bonitz},\ and\ \citenamefont {Ramazanov}}]{moldabekov2015}%
  \BibitemOpen
  \bibfield  {author} {\bibinfo {author} {\bibfnamefont {Zhandos}\ \bibnamefont {Moldabekov}}, \bibinfo {author} {\bibfnamefont {Patrick}\ \bibnamefont {Ludwig}}, \bibinfo {author} {\bibfnamefont {Michael}\ \bibnamefont {Bonitz}}, \ and\ \bibinfo {author} {\bibfnamefont {Tlekkabul}\ \bibnamefont {Ramazanov}},\ }\bibfield  {title} {\enquote {\bibinfo {title} {Ion potential in warm dense matter: Wake effects due to streaming degenerate electrons},}\ }\href@noop {} {\bibfield  {journal} {\bibinfo  {journal} {Physical Review E}\ }\textbf {\bibinfo {volume} {91}},\ \bibinfo {pages} {023102} (\bibinfo {year} {2015})}\BibitemShut {NoStop}%
\bibitem [{\citenamefont {Remington}\ \emph {et~al.}(2000)\citenamefont {Remington}, \citenamefont {Drake}, \citenamefont {Takabe},\ and\ \citenamefont {Arnett}}]{Remington2000}%
  \BibitemOpen
  \bibfield  {author} {\bibinfo {author} {\bibfnamefont {Bruce~A.}\ \bibnamefont {Remington}}, \bibinfo {author} {\bibfnamefont {R.~Paul}\ \bibnamefont {Drake}}, \bibinfo {author} {\bibfnamefont {Hideaki}\ \bibnamefont {Takabe}}, \ and\ \bibinfo {author} {\bibfnamefont {David}\ \bibnamefont {Arnett}},\ }\bibfield  {title} {\enquote {\bibinfo {title} {A review of astrophysics experiments on intense lasers},}\ }\href {\doibase 10.1063/1.874046} {\bibfield  {journal} {\bibinfo  {journal} {Physics of Plasmas}\ }\textbf {\bibinfo {volume} {7}},\ \bibinfo {pages} {1641--1652} (\bibinfo {year} {2000})},\ \Eprint {http://arxiv.org/abs/https://doi.org/10.1063/1.874046} {https://doi.org/10.1063/1.874046} \BibitemShut {NoStop}%
\bibitem [{\citenamefont {Glenzer}\ \emph {et~al.}(2010)\citenamefont {Glenzer}, \citenamefont {MacGowan}, \citenamefont {Michel}, \citenamefont {Meezan}, \citenamefont {Suter}, \citenamefont {Dixit}, \citenamefont {Kline}, \citenamefont {Kyrala}, \citenamefont {Bradley}, \citenamefont {Callahan}, \citenamefont {Dewald}, \citenamefont {Divol}, \citenamefont {Dzenitis}, \citenamefont {Edwards}, \citenamefont {Hamza}, \citenamefont {Haynam}, \citenamefont {Hinkel}, \citenamefont {Kalantar}, \citenamefont {Kilkenny}, \citenamefont {Landen}, \citenamefont {Lindl}, \citenamefont {LePape}, \citenamefont {Moody}, \citenamefont {Nikroo}, \citenamefont {Parham}, \citenamefont {Schneider}, \citenamefont {Town}, \citenamefont {Wegner}, \citenamefont {Widmann}, \citenamefont {Whitman}, \citenamefont {Young}, \citenamefont {Wonterghem}, \citenamefont {Atherton},\ and\ \citenamefont {Moses}}]{Glenzer2010}%
  \BibitemOpen
  \bibfield  {author} {\bibinfo {author} {\bibfnamefont {S.~H.}\ \bibnamefont {Glenzer}}, \bibinfo {author} {\bibfnamefont {B.~J.}\ \bibnamefont {MacGowan}}, \bibinfo {author} {\bibfnamefont {P.}~\bibnamefont {Michel}}, \bibinfo {author} {\bibfnamefont {N.~B.}\ \bibnamefont {Meezan}}, \bibinfo {author} {\bibfnamefont {L.~J.}\ \bibnamefont {Suter}}, \bibinfo {author} {\bibfnamefont {S.~N.}\ \bibnamefont {Dixit}}, \bibinfo {author} {\bibfnamefont {J.~L.}\ \bibnamefont {Kline}}, \bibinfo {author} {\bibfnamefont {G.~A.}\ \bibnamefont {Kyrala}}, \bibinfo {author} {\bibfnamefont {D.~K.}\ \bibnamefont {Bradley}}, \bibinfo {author} {\bibfnamefont {D.~A.}\ \bibnamefont {Callahan}}, \bibinfo {author} {\bibfnamefont {E.~L.}\ \bibnamefont {Dewald}}, \bibinfo {author} {\bibfnamefont {L.}~\bibnamefont {Divol}}, \bibinfo {author} {\bibfnamefont {E.}~\bibnamefont {Dzenitis}}, \bibinfo {author} {\bibfnamefont {M.~J.}\ \bibnamefont {Edwards}}, \bibinfo {author} {\bibfnamefont {A.~V.}\ \bibnamefont {Hamza}}, \bibinfo {author}
  {\bibfnamefont {C.~A.}\ \bibnamefont {Haynam}}, \bibinfo {author} {\bibfnamefont {D.~E.}\ \bibnamefont {Hinkel}}, \bibinfo {author} {\bibfnamefont {D.~H.}\ \bibnamefont {Kalantar}}, \bibinfo {author} {\bibfnamefont {J.~D.}\ \bibnamefont {Kilkenny}}, \bibinfo {author} {\bibfnamefont {O.~L.}\ \bibnamefont {Landen}}, \bibinfo {author} {\bibfnamefont {J.~D.}\ \bibnamefont {Lindl}}, \bibinfo {author} {\bibfnamefont {S.}~\bibnamefont {LePape}}, \bibinfo {author} {\bibfnamefont {J.~D.}\ \bibnamefont {Moody}}, \bibinfo {author} {\bibfnamefont {A.}~\bibnamefont {Nikroo}}, \bibinfo {author} {\bibfnamefont {T.}~\bibnamefont {Parham}}, \bibinfo {author} {\bibfnamefont {M.~B.}\ \bibnamefont {Schneider}}, \bibinfo {author} {\bibfnamefont {R.~P.~J.}\ \bibnamefont {Town}}, \bibinfo {author} {\bibfnamefont {P.}~\bibnamefont {Wegner}}, \bibinfo {author} {\bibfnamefont {K.}~\bibnamefont {Widmann}}, \bibinfo {author} {\bibfnamefont {P.}~\bibnamefont {Whitman}}, \bibinfo {author} {\bibfnamefont {B.~K.~F.}\ \bibnamefont
  {Young}}, \bibinfo {author} {\bibfnamefont {B.~Van}\ \bibnamefont {Wonterghem}}, \bibinfo {author} {\bibfnamefont {L.~J.}\ \bibnamefont {Atherton}}, \ and\ \bibinfo {author} {\bibfnamefont {E.~I.}\ \bibnamefont {Moses}},\ }\bibfield  {title} {\enquote {\bibinfo {title} {Symmetric inertial confinement fusion implosions at ultra-high laser energies},}\ }\href {\doibase 10.1126/science.1185634} {\bibfield  {journal} {\bibinfo  {journal} {Science}\ }\textbf {\bibinfo {volume} {327}},\ \bibinfo {pages} {1228--1231} (\bibinfo {year} {2010})}\BibitemShut {NoStop}%
\bibitem [{\citenamefont {Atzeni}\ and\ \citenamefont {Meyer-ter Vehn}(2004)}]{atzeni2004}%
  \BibitemOpen
  \bibfield  {author} {\bibinfo {author} {\bibfnamefont {S.}~\bibnamefont {Atzeni}}\ and\ \bibinfo {author} {\bibfnamefont {J.}~\bibnamefont {Meyer-ter Vehn}},\ }\href {https://books.google.de/books?id=BJcy\_p5pUBsC} {\emph {\bibinfo {title} {The Physics of Inertial Fusion: Beam Plasma Interaction, Hydrodynamics, Hot Dense Matter}}},\ International Series of Monographs on Physics\ (\bibinfo  {publisher} {OUP Oxford},\ \bibinfo {year} {2004})\BibitemShut {NoStop}%
\bibitem [{\citenamefont {Bonse}\ and\ \citenamefont {Gräf}(2020)}]{Bonse2015}%
  \BibitemOpen
  \bibfield  {author} {\bibinfo {author} {\bibfnamefont {Jörn}\ \bibnamefont {Bonse}}\ and\ \bibinfo {author} {\bibfnamefont {Stephan}\ \bibnamefont {Gräf}},\ }\bibfield  {title} {\enquote {\bibinfo {title} {Maxwell meets marangoni—a review of theories on laser-induced periodic surface structures},}\ }\href {\doibase 10.1002/lpor.202000215} {\bibfield  {journal} {\bibinfo  {journal} {Laser \& Photonics Reviews}\ }\textbf {\bibinfo {volume} {14}},\ \bibinfo {pages} {2000215} (\bibinfo {year} {2020})}\BibitemShut {NoStop}%
\bibitem [{\citenamefont {Nolte}\ \emph {et~al.}(1997)\citenamefont {Nolte}, \citenamefont {Momma}, \citenamefont {Jacobs}, \citenamefont {T{\"{u}}nnermann}, \citenamefont {Chichkov}, \citenamefont {Wellegehausen},\ and\ \citenamefont {Welling}}]{Nolte1997}%
  \BibitemOpen
  \bibfield  {author} {\bibinfo {author} {\bibfnamefont {S.}~\bibnamefont {Nolte}}, \bibinfo {author} {\bibfnamefont {C.}~\bibnamefont {Momma}}, \bibinfo {author} {\bibfnamefont {H.}~\bibnamefont {Jacobs}}, \bibinfo {author} {\bibfnamefont {A.}~\bibnamefont {T{\"{u}}nnermann}}, \bibinfo {author} {\bibfnamefont {B.~N.}\ \bibnamefont {Chichkov}}, \bibinfo {author} {\bibfnamefont {B.}~\bibnamefont {Wellegehausen}}, \ and\ \bibinfo {author} {\bibfnamefont {H.}~\bibnamefont {Welling}},\ }\bibfield  {title} {\enquote {\bibinfo {title} {{Ablation of metals by ultrashort laser pulses}},}\ }\href {\doibase 10.1364/JOSAB.14.002716} {\bibfield  {journal} {\bibinfo  {journal} {Journal of the Optical Society of America B}\ }\textbf {\bibinfo {volume} {14}},\ \bibinfo {pages} {2716} (\bibinfo {year} {1997})}\BibitemShut {NoStop}%
\bibitem [{\citenamefont {Sentoku}\ and\ \citenamefont {Kemp}(2008)}]{SENTOKU2008}%
  \BibitemOpen
  \bibfield  {author} {\bibinfo {author} {\bibfnamefont {Y.}~\bibnamefont {Sentoku}}\ and\ \bibinfo {author} {\bibfnamefont {A.J.}\ \bibnamefont {Kemp}},\ }\bibfield  {title} {\enquote {\bibinfo {title} {Numerical methods for particle simulations at extreme densities and temperatures: Weighted particles, relativistic collisions and reduced currents},}\ }\href {\doibase https://doi.org/10.1016/j.jcp.2008.03.043} {\bibfield  {journal} {\bibinfo  {journal} {Journal of Computational Physics}\ }\textbf {\bibinfo {volume} {227}},\ \bibinfo {pages} {6846 -- 6861} (\bibinfo {year} {2008})}\BibitemShut {NoStop}%
\bibitem [{\citenamefont {Antici}\ \emph {et~al.}(2008)\citenamefont {Antici}, \citenamefont {Fuchs}, \citenamefont {Borghesi}, \citenamefont {Gremillet}, \citenamefont {Grismayer}, \citenamefont {Sentoku}, \citenamefont {d'Humi\`eres}, \citenamefont {Cecchetti}, \citenamefont {Man\ifmmode \check{c}\else \v{c}\fi{}i\ifmmode~\acute{c}\else \'{c}\fi{}}, \citenamefont {Pipahl}, \citenamefont {Toncian}, \citenamefont {Willi}, \citenamefont {Mora},\ and\ \citenamefont {Audebert}}]{Antici2008}%
  \BibitemOpen
  \bibfield  {author} {\bibinfo {author} {\bibfnamefont {P.}~\bibnamefont {Antici}}, \bibinfo {author} {\bibfnamefont {J.}~\bibnamefont {Fuchs}}, \bibinfo {author} {\bibfnamefont {M.}~\bibnamefont {Borghesi}}, \bibinfo {author} {\bibfnamefont {L.}~\bibnamefont {Gremillet}}, \bibinfo {author} {\bibfnamefont {T.}~\bibnamefont {Grismayer}}, \bibinfo {author} {\bibfnamefont {Y.}~\bibnamefont {Sentoku}}, \bibinfo {author} {\bibfnamefont {E.}~\bibnamefont {d'Humi\`eres}}, \bibinfo {author} {\bibfnamefont {C.~A.}\ \bibnamefont {Cecchetti}}, \bibinfo {author} {\bibfnamefont {A.}~\bibnamefont {Man\ifmmode \check{c}\else \v{c}\fi{}i\ifmmode~\acute{c}\else \'{c}\fi{}}}, \bibinfo {author} {\bibfnamefont {A.~C.}\ \bibnamefont {Pipahl}}, \bibinfo {author} {\bibfnamefont {T.}~\bibnamefont {Toncian}}, \bibinfo {author} {\bibfnamefont {O.}~\bibnamefont {Willi}}, \bibinfo {author} {\bibfnamefont {P.}~\bibnamefont {Mora}}, \ and\ \bibinfo {author} {\bibfnamefont {P.}~\bibnamefont {Audebert}},\ }\bibfield  {title} {\enquote
  {\bibinfo {title} {Hot and cold electron dynamics following high-intensity laser matter interaction},}\ }\href {\doibase 10.1103/PhysRevLett.101.105004} {\bibfield  {journal} {\bibinfo  {journal} {Phys. Rev. Lett.}\ }\textbf {\bibinfo {volume} {101}},\ \bibinfo {pages} {105004} (\bibinfo {year} {2008})}\BibitemShut {NoStop}%
\bibitem [{\citenamefont {Kluge}\ \emph {et~al.}(2016)\citenamefont {Kluge}, \citenamefont {Bussmann}, \citenamefont {Chung}, \citenamefont {Gutt}, \citenamefont {Huang}, \citenamefont {Zacharias}, \citenamefont {Schramm},\ and\ \citenamefont {Cowan}}]{Kluge2016}%
  \BibitemOpen
  \bibfield  {author} {\bibinfo {author} {\bibfnamefont {T.}~\bibnamefont {Kluge}}, \bibinfo {author} {\bibfnamefont {M.}~\bibnamefont {Bussmann}}, \bibinfo {author} {\bibfnamefont {H.-K.}\ \bibnamefont {Chung}}, \bibinfo {author} {\bibfnamefont {C.}~\bibnamefont {Gutt}}, \bibinfo {author} {\bibfnamefont {L.~G.}\ \bibnamefont {Huang}}, \bibinfo {author} {\bibfnamefont {M.}~\bibnamefont {Zacharias}}, \bibinfo {author} {\bibfnamefont {U.}~\bibnamefont {Schramm}}, \ and\ \bibinfo {author} {\bibfnamefont {T.~E.}\ \bibnamefont {Cowan}},\ }\bibfield  {title} {\enquote {\bibinfo {title} {Nanoscale femtosecond imaging of transient hot solid density plasmas with elemental and charge state sensitivity using resonant coherent diffraction},}\ }\href {\doibase 10.1063/1.4942786} {\bibfield  {journal} {\bibinfo  {journal} {Physics of Plasmas}\ }\textbf {\bibinfo {volume} {23}},\ \bibinfo {pages} {033103} (\bibinfo {year} {2016})},\ \Eprint {http://arxiv.org/abs/https://doi.org/10.1063/1.4942786}
  {https://doi.org/10.1063/1.4942786} \BibitemShut {NoStop}%
\bibitem [{\citenamefont {Macchi}(2018)}]{macchi2018}%
  \BibitemOpen
  \bibfield  {author} {\bibinfo {author} {\bibfnamefont {A}~\bibnamefont {Macchi}},\ }\bibfield  {title} {\enquote {\bibinfo {title} {Surface plasmons in superintense laser-solid interactions},}\ }\href@noop {} {\bibfield  {journal} {\bibinfo  {journal} {Physics of Plasmas}\ }\textbf {\bibinfo {volume} {25}},\ \bibinfo {pages} {031906} (\bibinfo {year} {2018})}\BibitemShut {NoStop}%
\bibitem [{\citenamefont {Huang}\ \emph {et~al.}(2016)\citenamefont {Huang}, \citenamefont {Kluge},\ and\ \citenamefont {Cowan}}]{Huang2016}%
  \BibitemOpen
  \bibfield  {author} {\bibinfo {author} {\bibfnamefont {L.~G.}\ \bibnamefont {Huang}}, \bibinfo {author} {\bibfnamefont {T.}~\bibnamefont {Kluge}}, \ and\ \bibinfo {author} {\bibfnamefont {T.~E.}\ \bibnamefont {Cowan}},\ }\bibfield  {title} {\enquote {\bibinfo {title} {Dynamics of bulk electron heating and ionization in solid density plasmas driven by ultra-short relativistic laser pulses},}\ }\href {\doibase 10.1063/1.4953891} {\bibfield  {journal} {\bibinfo  {journal} {Physics of Plasmas}\ }\textbf {\bibinfo {volume} {23}},\ \bibinfo {pages} {063112} (\bibinfo {year} {2016})},\ \Eprint {http://arxiv.org/abs/https://doi.org/10.1063/1.4953891} {https://doi.org/10.1063/1.4953891} \BibitemShut {NoStop}%
\bibitem [{\citenamefont {Birdsall}\ and\ \citenamefont {Langdon}(2018)}]{birdsall2018}%
  \BibitemOpen
  \bibfield  {author} {\bibinfo {author} {\bibfnamefont {Charles~K}\ \bibnamefont {Birdsall}}\ and\ \bibinfo {author} {\bibfnamefont {A~Bruce}\ \bibnamefont {Langdon}},\ }\href@noop {} {\emph {\bibinfo {title} {Plasma physics via computer simulation}}}\ (\bibinfo  {publisher} {CRC press},\ \bibinfo {year} {2018})\BibitemShut {NoStop}%
\bibitem [{\citenamefont {Takizuka}\ and\ \citenamefont {Abe}(1977)}]{TA77}%
  \BibitemOpen
  \bibfield  {author} {\bibinfo {author} {\bibfnamefont {Tomonor}\ \bibnamefont {Takizuka}}\ and\ \bibinfo {author} {\bibfnamefont {Hirotada}\ \bibnamefont {Abe}},\ }\bibfield  {title} {\enquote {\bibinfo {title} {A binary collision model for plasma simulation with a particle code},}\ }\href {\doibase https://doi.org/10.1016/0021-9991(77)90099-7} {\bibfield  {journal} {\bibinfo  {journal} {Journal of Computational Physics}\ }\textbf {\bibinfo {volume} {25}},\ \bibinfo {pages} {205--219} (\bibinfo {year} {1977})}\BibitemShut {NoStop}%
\bibitem [{\citenamefont {Yang}\ \emph {et~al.}(2023)\citenamefont {Yang}, \citenamefont {Huang}, \citenamefont {Assenbaum}, \citenamefont {Cowan}, \citenamefont {Goethel}, \citenamefont {Göde}, \citenamefont {Kluge}, \citenamefont {Rehwald}, \citenamefont {Pan}, \citenamefont {Schramm}, \citenamefont {Vorberger}, \citenamefont {Zeil}, \citenamefont {Ziegler},\ and\ \citenamefont {Bernert}}]{yang2023}%
  \BibitemOpen
  \bibfield  {author} {\bibinfo {author} {\bibfnamefont {Long}\ \bibnamefont {Yang}}, \bibinfo {author} {\bibfnamefont {Lingen}\ \bibnamefont {Huang}}, \bibinfo {author} {\bibfnamefont {Stefan}\ \bibnamefont {Assenbaum}}, \bibinfo {author} {\bibfnamefont {Thomas~E.}\ \bibnamefont {Cowan}}, \bibinfo {author} {\bibfnamefont {Ilja}\ \bibnamefont {Goethel}}, \bibinfo {author} {\bibfnamefont {Sebastian}\ \bibnamefont {Göde}}, \bibinfo {author} {\bibfnamefont {Thomas}\ \bibnamefont {Kluge}}, \bibinfo {author} {\bibfnamefont {Martin}\ \bibnamefont {Rehwald}}, \bibinfo {author} {\bibfnamefont {Xiayun}\ \bibnamefont {Pan}}, \bibinfo {author} {\bibfnamefont {Ulrich}\ \bibnamefont {Schramm}}, \bibinfo {author} {\bibfnamefont {Jan}\ \bibnamefont {Vorberger}}, \bibinfo {author} {\bibfnamefont {Karl}\ \bibnamefont {Zeil}}, \bibinfo {author} {\bibfnamefont {Tim}\ \bibnamefont {Ziegler}}, \ and\ \bibinfo {author} {\bibfnamefont {Constantin}\ \bibnamefont {Bernert}},\ }\bibfield  {title} {\enquote {\bibinfo {title}
  {Time-resolved optical shadowgraphy of solid hydrogen jets as a testbed to benchmark particle-in-cell simulations},}\ }\href {\doibase 10.1038/s42005-023-01473-w} {\bibfield  {journal} {\bibinfo  {journal} {Communications Physics}\ }\textbf {\bibinfo {volume} {6}},\ \bibinfo {pages} {368} (\bibinfo {year} {2023})}\BibitemShut {NoStop}%
\bibitem [{\citenamefont {Rightley}\ and\ \citenamefont {Baalrud}(2021{\natexlab{b}})}]{rightley2021kinetic}%
  \BibitemOpen
  \bibfield  {author} {\bibinfo {author} {\bibfnamefont {Shane}\ \bibnamefont {Rightley}}\ and\ \bibinfo {author} {\bibfnamefont {Scott~D}\ \bibnamefont {Baalrud}},\ }\bibfield  {title} {\enquote {\bibinfo {title} {Kinetic model for electron-ion transport in warm dense matter},}\ }\href@noop {} {\bibfield  {journal} {\bibinfo  {journal} {Physical Review E}\ }\textbf {\bibinfo {volume} {103}},\ \bibinfo {pages} {063206} (\bibinfo {year} {2021}{\natexlab{b}})}\BibitemShut {NoStop}%
\bibitem [{\citenamefont {Starrett}(2018)}]{starrett2018coulomb}%
  \BibitemOpen
  \bibfield  {author} {\bibinfo {author} {\bibfnamefont {Charles~Edward}\ \bibnamefont {Starrett}},\ }\bibfield  {title} {\enquote {\bibinfo {title} {Coulomb log for conductivity of dense plasmas},}\ }\href@noop {} {\bibfield  {journal} {\bibinfo  {journal} {Physics of Plasmas}\ }\textbf {\bibinfo {volume} {25}},\ \bibinfo {pages} {092707} (\bibinfo {year} {2018})}\BibitemShut {NoStop}%
\bibitem [{\citenamefont {Filippov}\ \emph {et~al.}(2018)\citenamefont {Filippov}, \citenamefont {Starostin},\ and\ \citenamefont {Gryaznov}}]{filippov2018coulomb}%
  \BibitemOpen
  \bibfield  {author} {\bibinfo {author} {\bibfnamefont {AV}~\bibnamefont {Filippov}}, \bibinfo {author} {\bibfnamefont {AN}~\bibnamefont {Starostin}}, \ and\ \bibinfo {author} {\bibfnamefont {VK}~\bibnamefont {Gryaznov}},\ }\bibfield  {title} {\enquote {\bibinfo {title} {Coulomb logarithm in nonideal and degenerate plasmas.}}\ }\href@noop {} {\bibfield  {journal} {\bibinfo  {journal} {Journal of Experimental \& Theoretical Physics}\ }\textbf {\bibinfo {volume} {126}} (\bibinfo {year} {2018})}\BibitemShut {NoStop}%
\bibitem [{\citenamefont {Daligault}(2016)}]{daligault2016quantum}%
  \BibitemOpen
  \bibfield  {author} {\bibinfo {author} {\bibfnamefont {J{\'e}r{\^o}me}\ \bibnamefont {Daligault}},\ }\bibfield  {title} {\enquote {\bibinfo {title} {On the quantum landau collision operator and electron collisions in dense plasmas},}\ }\href@noop {} {\bibfield  {journal} {\bibinfo  {journal} {Physics of Plasmas}\ }\textbf {\bibinfo {volume} {23}},\ \bibinfo {pages} {032706} (\bibinfo {year} {2016})}\BibitemShut {NoStop}%
\bibitem [{\citenamefont {Price}\ \emph {et~al.}(1995)\citenamefont {Price}, \citenamefont {More}, \citenamefont {Walling}, \citenamefont {Guethlein}, \citenamefont {Shepherd}, \citenamefont {Stewart},\ and\ \citenamefont {White}}]{Price1995}%
  \BibitemOpen
  \bibfield  {author} {\bibinfo {author} {\bibfnamefont {D.~F.}\ \bibnamefont {Price}}, \bibinfo {author} {\bibfnamefont {R.~M.}\ \bibnamefont {More}}, \bibinfo {author} {\bibfnamefont {R.~S.}\ \bibnamefont {Walling}}, \bibinfo {author} {\bibfnamefont {G.}~\bibnamefont {Guethlein}}, \bibinfo {author} {\bibfnamefont {R.~L.}\ \bibnamefont {Shepherd}}, \bibinfo {author} {\bibfnamefont {R.~E.}\ \bibnamefont {Stewart}}, \ and\ \bibinfo {author} {\bibfnamefont {W.~E.}\ \bibnamefont {White}},\ }\bibfield  {title} {\enquote {\bibinfo {title} {Absorption of ultrashort laser pulses by solid targets heated rapidly to temperatures 1--1000 ev},}\ }\href {\doibase 10.1103/PhysRevLett.75.252} {\bibfield  {journal} {\bibinfo  {journal} {Phys. Rev. Lett.}\ }\textbf {\bibinfo {volume} {75}},\ \bibinfo {pages} {252--255} (\bibinfo {year} {1995})}\BibitemShut {NoStop}%
\bibitem [{\citenamefont {Eidmann}\ \emph {et~al.}(2000)\citenamefont {Eidmann}, \citenamefont {Meyer-ter Vehn}, \citenamefont {Schlegel},\ and\ \citenamefont {H\"uller}}]{Eidmann00}%
  \BibitemOpen
  \bibfield  {author} {\bibinfo {author} {\bibfnamefont {K.}~\bibnamefont {Eidmann}}, \bibinfo {author} {\bibfnamefont {J.}~\bibnamefont {Meyer-ter Vehn}}, \bibinfo {author} {\bibfnamefont {T.}~\bibnamefont {Schlegel}}, \ and\ \bibinfo {author} {\bibfnamefont {S.}~\bibnamefont {H\"uller}},\ }\bibfield  {title} {\enquote {\bibinfo {title} {Hydrodynamic simulation of subpicosecond laser interaction with solid-density matter},}\ }\href {\doibase 10.1103/PhysRevE.62.1202} {\bibfield  {journal} {\bibinfo  {journal} {Phys. Rev. E}\ }\textbf {\bibinfo {volume} {62}},\ \bibinfo {pages} {1202--1214} (\bibinfo {year} {2000})}\BibitemShut {NoStop}%
\bibitem [{\citenamefont {Vorberger}\ and\ \citenamefont {Gericke}(2014)}]{VORBERGER2014}%
  \BibitemOpen
  \bibfield  {author} {\bibinfo {author} {\bibfnamefont {J.}~\bibnamefont {Vorberger}}\ and\ \bibinfo {author} {\bibfnamefont {D.O.}\ \bibnamefont {Gericke}},\ }\bibfield  {title} {\enquote {\bibinfo {title} {Comparison of electron–ion energy transfer in dense plasmas obtained from numerical simulations and quantum kinetic theory},}\ }\href {\doibase https://doi.org/10.1016/j.hedp.2013.10.006} {\bibfield  {journal} {\bibinfo  {journal} {High Energy Density Physics}\ }\textbf {\bibinfo {volume} {10}},\ \bibinfo {pages} {1--8} (\bibinfo {year} {2014})}\BibitemShut {NoStop}%
\bibitem [{\citenamefont {Eliezer}(2002)}]{eliezer}%
  \BibitemOpen
  \bibfield  {author} {\bibinfo {author} {\bibfnamefont {Shalom}\ \bibnamefont {Eliezer}},\ }\href@noop {} {\emph {\bibinfo {title} {The Interaction of High-Power Lasers with Plamsa}}},\ Series in Plasma Physics\ (\bibinfo  {publisher} {Institute of Physics Publishing},\ \bibinfo {year} {2002})\BibitemShut {NoStop}%
\bibitem [{\citenamefont {Fourment}\ \emph {et~al.}(2014)\citenamefont {Fourment}, \citenamefont {Deneuville}, \citenamefont {Descamps}, \citenamefont {Dorchies}, \citenamefont {Petit}, \citenamefont {Peyrusse}, \citenamefont {Holst},\ and\ \citenamefont {Recoules}}]{fourment2014}%
  \BibitemOpen
  \bibfield  {author} {\bibinfo {author} {\bibfnamefont {C}~\bibnamefont {Fourment}}, \bibinfo {author} {\bibfnamefont {F}~\bibnamefont {Deneuville}}, \bibinfo {author} {\bibfnamefont {D}~\bibnamefont {Descamps}}, \bibinfo {author} {\bibfnamefont {F}~\bibnamefont {Dorchies}}, \bibinfo {author} {\bibfnamefont {S}~\bibnamefont {Petit}}, \bibinfo {author} {\bibfnamefont {O}~\bibnamefont {Peyrusse}}, \bibinfo {author} {\bibfnamefont {B}~\bibnamefont {Holst}}, \ and\ \bibinfo {author} {\bibfnamefont {V}~\bibnamefont {Recoules}},\ }\bibfield  {title} {\enquote {\bibinfo {title} {Experimental determination of temperature-dependent electron-electron collision frequency in isochorically heated warm dense gold},}\ }\href@noop {} {\bibfield  {journal} {\bibinfo  {journal} {Physical Review B}\ }\textbf {\bibinfo {volume} {89}},\ \bibinfo {pages} {161110} (\bibinfo {year} {2014})}\BibitemShut {NoStop}%
\bibitem [{\citenamefont {Fisher}\ \emph {et~al.}(2001)\citenamefont {Fisher}, \citenamefont {Fraenkel}, \citenamefont {Henis}, \citenamefont {Moshe},\ and\ \citenamefont {Eliezer}}]{Fisher01}%
  \BibitemOpen
  \bibfield  {author} {\bibinfo {author} {\bibfnamefont {D.}~\bibnamefont {Fisher}}, \bibinfo {author} {\bibfnamefont {M.}~\bibnamefont {Fraenkel}}, \bibinfo {author} {\bibfnamefont {Z.}~\bibnamefont {Henis}}, \bibinfo {author} {\bibfnamefont {E.}~\bibnamefont {Moshe}}, \ and\ \bibinfo {author} {\bibfnamefont {S.}~\bibnamefont {Eliezer}},\ }\bibfield  {title} {\enquote {\bibinfo {title} {Interband and intraband (drude) contributions to femtosecond laser absorption in aluminum},}\ }\href {\doibase 10.1103/PhysRevE.65.016409} {\bibfield  {journal} {\bibinfo  {journal} {Phys. Rev. E}\ }\textbf {\bibinfo {volume} {65}},\ \bibinfo {pages} {016409} (\bibinfo {year} {2001})}\BibitemShut {NoStop}%
\bibitem [{\citenamefont {Mueller}\ and\ \citenamefont {Rethfeld}(2013)}]{mueller2013}%
  \BibitemOpen
  \bibfield  {author} {\bibinfo {author} {\bibfnamefont {BY}~\bibnamefont {Mueller}}\ and\ \bibinfo {author} {\bibfnamefont {B}~\bibnamefont {Rethfeld}},\ }\bibfield  {title} {\enquote {\bibinfo {title} {Relaxation dynamics in laser-excited metals under nonequilibrium conditions},}\ }\href@noop {} {\bibfield  {journal} {\bibinfo  {journal} {Physical Review B}\ }\textbf {\bibinfo {volume} {87}},\ \bibinfo {pages} {035139} (\bibinfo {year} {2013})}\BibitemShut {NoStop}%
\bibitem [{\citenamefont {Meyer-ter Vehn}\ and\ \citenamefont {Ramis}(2019)}]{meyer2019}%
  \BibitemOpen
  \bibfield  {author} {\bibinfo {author} {\bibfnamefont {J{\"u}rgen}\ \bibnamefont {Meyer-ter Vehn}}\ and\ \bibinfo {author} {\bibfnamefont {Rafael}\ \bibnamefont {Ramis}},\ }\bibfield  {title} {\enquote {\bibinfo {title} {On collisional free-free photon absorption in warm dense matter},}\ }\href@noop {} {\bibfield  {journal} {\bibinfo  {journal} {Physics of Plasmas}\ }\textbf {\bibinfo {volume} {26}},\ \bibinfo {pages} {113301} (\bibinfo {year} {2019})}\BibitemShut {NoStop}%
\bibitem [{\citenamefont {Lugovskoy}\ and\ \citenamefont {Bray}(1999)}]{lugovskoy1999}%
  \BibitemOpen
  \bibfield  {author} {\bibinfo {author} {\bibfnamefont {Andrey~V}\ \bibnamefont {Lugovskoy}}\ and\ \bibinfo {author} {\bibfnamefont {Igor}\ \bibnamefont {Bray}},\ }\bibfield  {title} {\enquote {\bibinfo {title} {Ultrafast electron dynamics in metals under laser irradiation},}\ }\href@noop {} {\bibfield  {journal} {\bibinfo  {journal} {Physical Review B}\ }\textbf {\bibinfo {volume} {60}},\ \bibinfo {pages} {3279} (\bibinfo {year} {1999})}\BibitemShut {NoStop}%
\bibitem [{\citenamefont {Petrov}\ \emph {et~al.}(2013)\citenamefont {Petrov}, \citenamefont {Inogamov},\ and\ \citenamefont {Migdal}}]{petrov2013}%
  \BibitemOpen
  \bibfield  {author} {\bibinfo {author} {\bibfnamefont {Yu~V}\ \bibnamefont {Petrov}}, \bibinfo {author} {\bibfnamefont {Nail'Alimovich}\ \bibnamefont {Inogamov}}, \ and\ \bibinfo {author} {\bibfnamefont {Kiril~Petrovich}\ \bibnamefont {Migdal}},\ }\bibfield  {title} {\enquote {\bibinfo {title} {Thermal conductivity and the electron-ion heat transfer coefficient in condensed media with a strongly excited electron subsystem},}\ }\href@noop {} {\bibfield  {journal} {\bibinfo  {journal} {JETP letters}\ }\textbf {\bibinfo {volume} {97}},\ \bibinfo {pages} {20--27} (\bibinfo {year} {2013})}\BibitemShut {NoStop}%
\bibitem [{\citenamefont {Pineau}\ \emph {et~al.}(2020)\citenamefont {Pineau}, \citenamefont {Chimier}, \citenamefont {Hu},\ and\ \citenamefont {Duchateau}}]{pineau2020modeling}%
  \BibitemOpen
  \bibfield  {author} {\bibinfo {author} {\bibfnamefont {A}~\bibnamefont {Pineau}}, \bibinfo {author} {\bibfnamefont {B}~\bibnamefont {Chimier}}, \bibinfo {author} {\bibfnamefont {SX}~\bibnamefont {Hu}}, \ and\ \bibinfo {author} {\bibfnamefont {G}~\bibnamefont {Duchateau}},\ }\bibfield  {title} {\enquote {\bibinfo {title} {Modeling the electron collision frequency during solid-to-plasma transition of polystyrene ablator for direct-drive inertial confinement fusion applications},}\ }\href@noop {} {\bibfield  {journal} {\bibinfo  {journal} {Physics of Plasmas}\ }\textbf {\bibinfo {volume} {27}} (\bibinfo {year} {2020})}\BibitemShut {NoStop}%
\bibitem [{\citenamefont {Ramis}\ \emph {et~al.}(2012)\citenamefont {Ramis}, \citenamefont {Eidmann}, \citenamefont {Meyer-ter Vehn},\ and\ \citenamefont {H{\"u}ller}}]{Ramis2012}%
  \BibitemOpen
  \bibfield  {author} {\bibinfo {author} {\bibfnamefont {R}~\bibnamefont {Ramis}}, \bibinfo {author} {\bibfnamefont {K}~\bibnamefont {Eidmann}}, \bibinfo {author} {\bibfnamefont {J}~\bibnamefont {Meyer-ter Vehn}}, \ and\ \bibinfo {author} {\bibfnamefont {S}~\bibnamefont {H{\"u}ller}},\ }\bibfield  {title} {\enquote {\bibinfo {title} {Multi-fs--a computer code for laser--plasma interaction in the femtosecond regime},}\ }\href@noop {} {\bibfield  {journal} {\bibinfo  {journal} {Computer Physics Communications}\ }\textbf {\bibinfo {volume} {183}},\ \bibinfo {pages} {637--655} (\bibinfo {year} {2012})}\BibitemShut {NoStop}%
\bibitem [{\citenamefont {Spitzer}(1956)}]{Spitzer}%
  \BibitemOpen
  \bibfield  {author} {\bibinfo {author} {\bibfnamefont {L.}~\bibnamefont {Spitzer}},\ }\href@noop {} {\emph {\bibinfo {title} {Physics of fully ionized gases}}}\ (\bibinfo  {publisher} {Interscience, New York},\ \bibinfo {year} {1956})\BibitemShut {NoStop}%
\bibitem [{\citenamefont {Lee}\ and\ \citenamefont {More}(1984)}]{Lee1984}%
  \BibitemOpen
  \bibfield  {author} {\bibinfo {author} {\bibfnamefont {Y.~T.}\ \bibnamefont {Lee}}\ and\ \bibinfo {author} {\bibfnamefont {R.~M.}\ \bibnamefont {More}},\ }\bibfield  {title} {\enquote {\bibinfo {title} {An electron conductivity model for dense plasmas},}\ }\href {\doibase 10.1063/1.864744} {\bibfield  {journal} {\bibinfo  {journal} {The Physics of Fluids}\ }\textbf {\bibinfo {volume} {27}},\ \bibinfo {pages} {1273--1286} (\bibinfo {year} {1984})},\ \Eprint {http://arxiv.org/abs/https://aip.scitation.org/doi/pdf/10.1063/1.864744} {https://aip.scitation.org/doi/pdf/10.1063/1.864744} \BibitemShut {NoStop}%
\bibitem [{\citenamefont {Brysk}\ \emph {et~al.}(1975)\citenamefont {Brysk}, \citenamefont {Campbell},\ and\ \citenamefont {Hammerling}}]{Brysk_1975}%
  \BibitemOpen
  \bibfield  {author} {\bibinfo {author} {\bibfnamefont {H}~\bibnamefont {Brysk}}, \bibinfo {author} {\bibfnamefont {P~M}\ \bibnamefont {Campbell}}, \ and\ \bibinfo {author} {\bibfnamefont {P}~\bibnamefont {Hammerling}},\ }\bibfield  {title} {\enquote {\bibinfo {title} {Thermal conduction in laser fusion},}\ }\href {\doibase 10.1088/0032-1028/17/6/007} {\bibfield  {journal} {\bibinfo  {journal} {Plasma Physics}\ }\textbf {\bibinfo {volume} {17}},\ \bibinfo {pages} {473--484} (\bibinfo {year} {1975})}\BibitemShut {NoStop}%
\bibitem [{\citenamefont {Kodanova}\ \emph {et~al.}(2018)\citenamefont {Kodanova}, \citenamefont {Issanova}, \citenamefont {Amirov}, \citenamefont {Ramazanov}, \citenamefont {Tikhonov},\ and\ \citenamefont {Moldabekov}}]{kodanova2018}%
  \BibitemOpen
  \bibfield  {author} {\bibinfo {author} {\bibfnamefont {SK}~\bibnamefont {Kodanova}}, \bibinfo {author} {\bibfnamefont {MK}~\bibnamefont {Issanova}}, \bibinfo {author} {\bibfnamefont {SM}~\bibnamefont {Amirov}}, \bibinfo {author} {\bibfnamefont {TS}~\bibnamefont {Ramazanov}}, \bibinfo {author} {\bibfnamefont {A}~\bibnamefont {Tikhonov}}, \ and\ \bibinfo {author} {\bibfnamefont {Zh~A}\ \bibnamefont {Moldabekov}},\ }\bibfield  {title} {\enquote {\bibinfo {title} {Relaxation of non-isothermal hot dense plasma parameters},}\ }\href@noop {} {\bibfield  {journal} {\bibinfo  {journal} {Matter and Radiation at Extremes}\ }\textbf {\bibinfo {volume} {3}},\ \bibinfo {pages} {40--49} (\bibinfo {year} {2018})}\BibitemShut {NoStop}%
\bibitem [{\citenamefont {Guethlein}\ \emph {et~al.}(1996)\citenamefont {Guethlein}, \citenamefont {Foord},\ and\ \citenamefont {Price}}]{guethlein1996}%
  \BibitemOpen
  \bibfield  {author} {\bibinfo {author} {\bibfnamefont {G}~\bibnamefont {Guethlein}}, \bibinfo {author} {\bibfnamefont {ME}~\bibnamefont {Foord}}, \ and\ \bibinfo {author} {\bibfnamefont {D}~\bibnamefont {Price}},\ }\bibfield  {title} {\enquote {\bibinfo {title} {Electron temperature measurements of solid density plasmas produced by intense ultrashort laser pulses},}\ }\href@noop {} {\bibfield  {journal} {\bibinfo  {journal} {Physical review letters}\ }\textbf {\bibinfo {volume} {77}},\ \bibinfo {pages} {1055} (\bibinfo {year} {1996})}\BibitemShut {NoStop}%
\bibitem [{\citenamefont {Randolph}\ \emph {et~al.}(2022)\citenamefont {Randolph}, \citenamefont {Banjafar}, \citenamefont {Preston}, \citenamefont {Yabuuchi}, \citenamefont {Makita}, \citenamefont {Dover}, \citenamefont {R\"odel}, \citenamefont {G\"ode}, \citenamefont {Inubushi}, \citenamefont {Jakob}, \citenamefont {Kaa}, \citenamefont {Kon}, \citenamefont {Koga}, \citenamefont {Ksenzov}, \citenamefont {Matsuoka}, \citenamefont {Nishiuchi}, \citenamefont {Paulus}, \citenamefont {Schon}, \citenamefont {Sueda}, \citenamefont {Sentoku}, \citenamefont {Togashi}, \citenamefont {Bussmann}, \citenamefont {Cowan}, \citenamefont {Kl\"aui}, \citenamefont {Fortmann-Grote}, \citenamefont {Huang}, \citenamefont {Mancuso}, \citenamefont {Kluge}, \citenamefont {Gutt},\ and\ \citenamefont {Nakatsutsumi}}]{randolph20}%
  \BibitemOpen
  \bibfield  {author} {\bibinfo {author} {\bibfnamefont {Lisa}\ \bibnamefont {Randolph}}, \bibinfo {author} {\bibfnamefont {Mohammadreza}\ \bibnamefont {Banjafar}}, \bibinfo {author} {\bibfnamefont {Thomas~R.}\ \bibnamefont {Preston}}, \bibinfo {author} {\bibfnamefont {Toshinori}\ \bibnamefont {Yabuuchi}}, \bibinfo {author} {\bibfnamefont {Mikako}\ \bibnamefont {Makita}}, \bibinfo {author} {\bibfnamefont {Nicholas~P.}\ \bibnamefont {Dover}}, \bibinfo {author} {\bibfnamefont {Christian}\ \bibnamefont {R\"odel}}, \bibinfo {author} {\bibfnamefont {Sebastian}\ \bibnamefont {G\"ode}}, \bibinfo {author} {\bibfnamefont {Yuichi}\ \bibnamefont {Inubushi}}, \bibinfo {author} {\bibfnamefont {Gerhard}\ \bibnamefont {Jakob}}, \bibinfo {author} {\bibfnamefont {Johannes}\ \bibnamefont {Kaa}}, \bibinfo {author} {\bibfnamefont {Akira}\ \bibnamefont {Kon}}, \bibinfo {author} {\bibfnamefont {James~K.}\ \bibnamefont {Koga}}, \bibinfo {author} {\bibfnamefont {Dmitriy}\ \bibnamefont {Ksenzov}}, \bibinfo {author} {\bibfnamefont
  {Takeshi}\ \bibnamefont {Matsuoka}}, \bibinfo {author} {\bibfnamefont {Mamiko}\ \bibnamefont {Nishiuchi}}, \bibinfo {author} {\bibfnamefont {Michael}\ \bibnamefont {Paulus}}, \bibinfo {author} {\bibfnamefont {Frederic}\ \bibnamefont {Schon}}, \bibinfo {author} {\bibfnamefont {Keiichi}\ \bibnamefont {Sueda}}, \bibinfo {author} {\bibfnamefont {Yasuhiko}\ \bibnamefont {Sentoku}}, \bibinfo {author} {\bibfnamefont {Tadashi}\ \bibnamefont {Togashi}}, \bibinfo {author} {\bibfnamefont {Michael}\ \bibnamefont {Bussmann}}, \bibinfo {author} {\bibfnamefont {Thomas~E.}\ \bibnamefont {Cowan}}, \bibinfo {author} {\bibfnamefont {Mathias}\ \bibnamefont {Kl\"aui}}, \bibinfo {author} {\bibfnamefont {Carsten}\ \bibnamefont {Fortmann-Grote}}, \bibinfo {author} {\bibfnamefont {Lingen}\ \bibnamefont {Huang}}, \bibinfo {author} {\bibfnamefont {Adrian~P.}\ \bibnamefont {Mancuso}}, \bibinfo {author} {\bibfnamefont {Thomas}\ \bibnamefont {Kluge}}, \bibinfo {author} {\bibfnamefont {Christian}\ \bibnamefont {Gutt}}, \ and\ \bibinfo
  {author} {\bibfnamefont {Motoaki}\ \bibnamefont {Nakatsutsumi}},\ }\bibfield  {title} {\enquote {\bibinfo {title} {Nanoscale subsurface dynamics of solids upon high-intensity femtosecond laser irradiation observed by grazing-incidence x-ray scattering},}\ }\href {\doibase 10.1103/PhysRevResearch.4.033038} {\bibfield  {journal} {\bibinfo  {journal} {Phys. Rev. Research}\ }\textbf {\bibinfo {volume} {4}},\ \bibinfo {pages} {033038} (\bibinfo {year} {2022})}\BibitemShut {NoStop}%
\bibitem [{\citenamefont {Randolph}(2020)}]{LisaThesis}%
  \BibitemOpen
  \bibfield  {author} {\bibinfo {author} {\bibfnamefont {Lisa}\ \bibnamefont {Randolph}},\ }\emph {\bibinfo {title} {Surface dynamics of solids upon high-intensity laser irradiation investigated by grazing incidence X-ray scattering}},\ \href {\doibase http://dx.doi.org/10.25819/ubsi/6596} {Ph.D. thesis},\ \bibinfo  {school} {Universität Siegen} (\bibinfo {year} {2020})\BibitemShut {NoStop}%
\bibitem [{\citenamefont {Yabuuchi}\ \emph {et~al.}(2019)\citenamefont {Yabuuchi}, \citenamefont {Kon}, \citenamefont {Inubushi}, \citenamefont {Togahi}, \citenamefont {Sueda}, \citenamefont {Itoga}, \citenamefont {Nakajima}, \citenamefont {Habara}, \citenamefont {Kodama}, \citenamefont {Tomizawa},\ and\ \citenamefont {Yabashi}}]{Yabuuchi19}%
  \BibitemOpen
  \bibfield  {author} {\bibinfo {author} {\bibfnamefont {T.}~\bibnamefont {Yabuuchi}}, \bibinfo {author} {\bibfnamefont {A.}~\bibnamefont {Kon}}, \bibinfo {author} {\bibfnamefont {Y.}~\bibnamefont {Inubushi}}, \bibinfo {author} {\bibfnamefont {T.}~\bibnamefont {Togahi}}, \bibinfo {author} {\bibfnamefont {K.}~\bibnamefont {Sueda}}, \bibinfo {author} {\bibfnamefont {T.}~\bibnamefont {Itoga}}, \bibinfo {author} {\bibfnamefont {K.}~\bibnamefont {Nakajima}}, \bibinfo {author} {\bibfnamefont {H.}~\bibnamefont {Habara}}, \bibinfo {author} {\bibfnamefont {R.}~\bibnamefont {Kodama}}, \bibinfo {author} {\bibfnamefont {H.}~\bibnamefont {Tomizawa}}, \ and\ \bibinfo {author} {\bibfnamefont {M.}~\bibnamefont {Yabashi}},\ }\bibfield  {title} {\enquote {\bibinfo {title} {{An experimental platform using high-power, high-intensity optical lasers with the hard X-ray free-electron laser at SACLA}},}\ }\href {\doibase 10.1107/S1600577519000882} {\bibfield  {journal} {\bibinfo  {journal} {J. Sync. Rad.}\ }\textbf {\bibinfo
  {volume} {26}},\ \bibinfo {pages} {585--594} (\bibinfo {year} {2019})}\BibitemShut {NoStop}%
\bibitem [{\citenamefont {Kameshima}\ \emph {et~al.}(2014)\citenamefont {Kameshima}, \citenamefont {Ono}, \citenamefont {Kudo}, \citenamefont {Ozaki}, \citenamefont {Kirihara}, \citenamefont {Kobayashi}, \citenamefont {Inubushi}, \citenamefont {Yabashi}, \citenamefont {Horigome}, \citenamefont {Holland}, \citenamefont {Holland}, \citenamefont {Burt}, \citenamefont {Murao},\ and\ \citenamefont {Hatsui}}]{Kameshima14}%
  \BibitemOpen
  \bibfield  {author} {\bibinfo {author} {\bibfnamefont {T.}~\bibnamefont {Kameshima}}, \bibinfo {author} {\bibfnamefont {S.}~\bibnamefont {Ono}}, \bibinfo {author} {\bibfnamefont {T.}~\bibnamefont {Kudo}}, \bibinfo {author} {\bibfnamefont {K.}~\bibnamefont {Ozaki}}, \bibinfo {author} {\bibfnamefont {Y.}~\bibnamefont {Kirihara}}, \bibinfo {author} {\bibfnamefont {K.}~\bibnamefont {Kobayashi}}, \bibinfo {author} {\bibfnamefont {Y.}~\bibnamefont {Inubushi}}, \bibinfo {author} {\bibfnamefont {M.}~\bibnamefont {Yabashi}}, \bibinfo {author} {\bibfnamefont {T.}~\bibnamefont {Horigome}}, \bibinfo {author} {\bibfnamefont {A.}~\bibnamefont {Holland}}, \bibinfo {author} {\bibfnamefont {K.}~\bibnamefont {Holland}}, \bibinfo {author} {\bibfnamefont {D.}~\bibnamefont {Burt}}, \bibinfo {author} {\bibfnamefont {H.}~\bibnamefont {Murao}}, \ and\ \bibinfo {author} {\bibfnamefont {T.}~\bibnamefont {Hatsui}},\ }\bibfield  {title} {\enquote {\bibinfo {title} {{Development of an X-ray pixel detector with multi-port charge-coupled
  device for X-ray free-electron laser experiments}},}\ }\href {\doibase 10.1063/1.4867668} {\bibfield  {journal} {\bibinfo  {journal} {Review of Scientific Instruments}\ }\textbf {\bibinfo {volume} {85}},\ \bibinfo {pages} {033110} (\bibinfo {year} {2014})}\BibitemShut {NoStop}%
\bibitem [{\citenamefont {Hol\'{y}}\ \emph {et~al.}(1993)\citenamefont {Hol\'{y}}, \citenamefont {Kub\v{e}na}, \citenamefont {Ohl\'{i}dal}, \citenamefont {Lischka},\ and\ \citenamefont {Plotz}}]{Holy93}%
  \BibitemOpen
  \bibfield  {author} {\bibinfo {author} {\bibfnamefont {V.}~\bibnamefont {Hol\'{y}}}, \bibinfo {author} {\bibfnamefont {J.}~\bibnamefont {Kub\v{e}na}}, \bibinfo {author} {\bibfnamefont {I.}~\bibnamefont {Ohl\'{i}dal}}, \bibinfo {author} {\bibfnamefont {K.}~\bibnamefont {Lischka}}, \ and\ \bibinfo {author} {\bibfnamefont {W.}~\bibnamefont {Plotz}},\ }\bibfield  {title} {\enquote {\bibinfo {title} {{X-ray reflection from rough layered systems}},}\ }\href {\doibase 10.1103/PhysRevB.47.15896} {\bibfield  {journal} {\bibinfo  {journal} {Phys. Rev. B}\ }\textbf {\bibinfo {volume} {47}},\ \bibinfo {pages} {15896--15903} (\bibinfo {year} {1993})}\BibitemShut {NoStop}%
\bibitem [{\citenamefont {Kiessig}(1931)}]{Kiessig31}%
  \BibitemOpen
  \bibfield  {author} {\bibinfo {author} {\bibfnamefont {H.}~\bibnamefont {Kiessig}},\ }\bibfield  {title} {\enquote {\bibinfo {title} {{Untersuchunger zur Totalreflexion von R\"{o}ntgenstrahlen}},}\ }\href@noop {} {\bibfield  {journal} {\bibinfo  {journal} {Annalen der Physik}\ }\textbf {\bibinfo {volume} {10}},\ \bibinfo {pages} {769} (\bibinfo {year} {1931})}\BibitemShut {NoStop}%
\bibitem [{\citenamefont {Yoneda}(1963)}]{Yoneda63}%
  \BibitemOpen
  \bibfield  {author} {\bibinfo {author} {\bibfnamefont {Y.}~\bibnamefont {Yoneda}},\ }\bibfield  {title} {\enquote {\bibinfo {title} {{Anomalous Surface Reflection of X Rays}},}\ }\href {\doibase 10.1103/PhysRev.131.2010} {\bibfield  {journal} {\bibinfo  {journal} {Phys. Rev.}\ }\textbf {\bibinfo {volume} {131}},\ \bibinfo {pages} {2010--2013} (\bibinfo {year} {1963})}\BibitemShut {NoStop}%
\bibitem [{\citenamefont {Chung}\ \emph {et~al.}(2005)\citenamefont {Chung}, \citenamefont {Chen}, \citenamefont {Morgan}, \citenamefont {Ralchenko},\ and\ \citenamefont {Lee}}]{Chung05}%
  \BibitemOpen
  \bibfield  {author} {\bibinfo {author} {\bibfnamefont {H.-K.}\ \bibnamefont {Chung}}, \bibinfo {author} {\bibfnamefont {M.H.}\ \bibnamefont {Chen}}, \bibinfo {author} {\bibfnamefont {W.L.}\ \bibnamefont {Morgan}}, \bibinfo {author} {\bibfnamefont {Y.}~\bibnamefont {Ralchenko}}, \ and\ \bibinfo {author} {\bibfnamefont {R.W.}\ \bibnamefont {Lee}},\ }\bibfield  {title} {\enquote {\bibinfo {title} {{FLYCHK: Generalized population kinetics and spectral model for rapid spectroscopic analysis for all elements}},}\ }\href {\doibase https://doi.org/10.1016/j.hedp.2005.07.001} {\bibfield  {journal} {\bibinfo  {journal} {High Energy Density Physics}\ }\textbf {\bibinfo {volume} {1}},\ \bibinfo {pages} {3 -- 12} (\bibinfo {year} {2005})}\BibitemShut {NoStop}%
\bibitem [{\citenamefont {Faik}\ \emph {et~al.}(2018)\citenamefont {Faik}, \citenamefont {Tauschwitz},\ and\ \citenamefont {Iosilevskiy}}]{Faik18}%
  \BibitemOpen
  \bibfield  {author} {\bibinfo {author} {\bibfnamefont {S.}~\bibnamefont {Faik}}, \bibinfo {author} {\bibfnamefont {A.}~\bibnamefont {Tauschwitz}}, \ and\ \bibinfo {author} {\bibfnamefont {I.}~\bibnamefont {Iosilevskiy}},\ }\bibfield  {title} {\enquote {\bibinfo {title} {{The equation of state package FEOS for high energy density matter}},}\ }\href {\doibase https://doi.org/10.1016/j.cpc.2018.01.008} {\bibfield  {journal} {\bibinfo  {journal} {Comput. Phys. Comm.}\ }\textbf {\bibinfo {volume} {227}},\ \bibinfo {pages} {117 -- 125} (\bibinfo {year} {2018})}\BibitemShut {NoStop}%
\bibitem [{\citenamefont {Drake}(2018)}]{Drake}%
  \BibitemOpen
  \bibfield  {author} {\bibinfo {author} {\bibfnamefont {R.~Paul}\ \bibnamefont {Drake}},\ }\href@noop {} {\emph {\bibinfo {title} {High-Energy-Density Physics: Fundamentals, Inertial Fusion, and Experimental Astrophysics}}}\ (\bibinfo  {publisher} {Springer, Cham},\ \bibinfo {year} {2018})\BibitemShut {NoStop}%
\bibitem [{\citenamefont {Dornheim}\ \emph {et~al.}(2023)\citenamefont {Dornheim}, \citenamefont {Moldabekov}, \citenamefont {Ramakrishna}, \citenamefont {Tolias}, \citenamefont {Baczewski}, \citenamefont {Kraus}, \citenamefont {Preston}, \citenamefont {Chapman}, \citenamefont {B{\"o}hme}, \citenamefont {D{\"o}ppner} \emph {et~al.}}]{dornheim2023}%
  \BibitemOpen
  \bibfield  {author} {\bibinfo {author} {\bibfnamefont {Tobias}\ \bibnamefont {Dornheim}}, \bibinfo {author} {\bibfnamefont {Zhandos~A}\ \bibnamefont {Moldabekov}}, \bibinfo {author} {\bibfnamefont {Kushal}\ \bibnamefont {Ramakrishna}}, \bibinfo {author} {\bibfnamefont {Panagiotis}\ \bibnamefont {Tolias}}, \bibinfo {author} {\bibfnamefont {Andrew~D}\ \bibnamefont {Baczewski}}, \bibinfo {author} {\bibfnamefont {Dominik}\ \bibnamefont {Kraus}}, \bibinfo {author} {\bibfnamefont {Thomas~R}\ \bibnamefont {Preston}}, \bibinfo {author} {\bibfnamefont {David~A}\ \bibnamefont {Chapman}}, \bibinfo {author} {\bibfnamefont {Maximilian~P}\ \bibnamefont {B{\"o}hme}}, \bibinfo {author} {\bibfnamefont {Tilo}\ \bibnamefont {D{\"o}ppner}},  \emph {et~al.},\ }\bibfield  {title} {\enquote {\bibinfo {title} {Electronic density response of warm dense matter},}\ }\href@noop {} {\bibfield  {journal} {\bibinfo  {journal} {Physics of Plasmas}\ }\textbf {\bibinfo {volume} {30}} (\bibinfo {year} {2023})}\BibitemShut {NoStop}%
\bibitem [{\citenamefont {Bussmann}\ \emph {et~al.}(2006)\citenamefont {Bussmann}, \citenamefont {Schramm},\ and\ \citenamefont {Habs}}]{bussmann2006}%
  \BibitemOpen
  \bibfield  {author} {\bibinfo {author} {\bibfnamefont {M}~\bibnamefont {Bussmann}}, \bibinfo {author} {\bibfnamefont {U}~\bibnamefont {Schramm}}, \ and\ \bibinfo {author} {\bibfnamefont {D}~\bibnamefont {Habs}},\ }\bibfield  {title} {\enquote {\bibinfo {title} {Simulating strongly coupled plasmas at low temperatures},}\ }in\ \href@noop {} {\emph {\bibinfo {booktitle} {AIP Conference Proceedings}}},\ Vol.\ \bibinfo {volume} {862}\ (\bibinfo {organization} {American Institute of Physics},\ \bibinfo {year} {2006})\ pp.\ \bibinfo {pages} {221--231}\BibitemShut {NoStop}%
\bibitem [{\citenamefont {Bussmann}(2008)}]{bussmann2008}%
  \BibitemOpen
  \bibfield  {author} {\bibinfo {author} {\bibfnamefont {Michael}\ \bibnamefont {Bussmann}},\ }\bibfield  {title} {\enquote {\bibinfo {title} {Laser cooled ion beams and strongly coupled plasmas for precision experiments},}\ }\href@noop {} {\  (\bibinfo {year} {2008})}\BibitemShut {NoStop}%
\bibitem [{\citenamefont {Peltz}\ \emph {et~al.}(2012)\citenamefont {Peltz}, \citenamefont {Varin}, \citenamefont {Brabec},\ and\ \citenamefont {Fennel}}]{peltz2012}%
  \BibitemOpen
  \bibfield  {author} {\bibinfo {author} {\bibfnamefont {Christian}\ \bibnamefont {Peltz}}, \bibinfo {author} {\bibfnamefont {Charles}\ \bibnamefont {Varin}}, \bibinfo {author} {\bibfnamefont {Thomas}\ \bibnamefont {Brabec}}, \ and\ \bibinfo {author} {\bibfnamefont {Thomas}\ \bibnamefont {Fennel}},\ }\bibfield  {title} {\enquote {\bibinfo {title} {Fully microscopic analysis of laser-driven finite plasmas using the example of clusters},}\ }\href@noop {} {\bibfield  {journal} {\bibinfo  {journal} {New Journal of Physics}\ }\textbf {\bibinfo {volume} {14}},\ \bibinfo {pages} {065011} (\bibinfo {year} {2012})}\BibitemShut {NoStop}%
\bibitem [{\citenamefont {H{\"u}bl}(2019)}]{hubl2019}%
  \BibitemOpen
  \bibfield  {author} {\bibinfo {author} {\bibfnamefont {Axel}\ \bibnamefont {H{\"u}bl}},\ }\bibfield  {title} {\enquote {\bibinfo {title} {Picongpu. predictive simulations of laser-particle accelerators with manycore hardware},}\ }\href {\doibase 10.5281/zenodo.3266820} {\  (\bibinfo {year} {2019}),\ 10.5281/zenodo.3266820}\BibitemShut {NoStop}%
\bibitem [{\citenamefont {Burau}\ \emph {et~al.}(2010)\citenamefont {Burau}, \citenamefont {Widera}, \citenamefont {Hönig}, \citenamefont {Juckeland}, \citenamefont {Debus}, \citenamefont {Kluge}, \citenamefont {Schramm}, \citenamefont {Cowan}, \citenamefont {Sauerbrey},\ and\ \citenamefont {Bussmann}}]{PIConGPU}%
  \BibitemOpen
  \bibfield  {author} {\bibinfo {author} {\bibfnamefont {Heiko}\ \bibnamefont {Burau}}, \bibinfo {author} {\bibfnamefont {Renée}\ \bibnamefont {Widera}}, \bibinfo {author} {\bibfnamefont {Wolfgang}\ \bibnamefont {Hönig}}, \bibinfo {author} {\bibfnamefont {Guido}\ \bibnamefont {Juckeland}}, \bibinfo {author} {\bibfnamefont {Alexander}\ \bibnamefont {Debus}}, \bibinfo {author} {\bibfnamefont {Thomas}\ \bibnamefont {Kluge}}, \bibinfo {author} {\bibfnamefont {Ulrich}\ \bibnamefont {Schramm}}, \bibinfo {author} {\bibfnamefont {Tomas~E.}\ \bibnamefont {Cowan}}, \bibinfo {author} {\bibfnamefont {Roland}\ \bibnamefont {Sauerbrey}}, \ and\ \bibinfo {author} {\bibfnamefont {Michael}\ \bibnamefont {Bussmann}},\ }\bibfield  {title} {\enquote {\bibinfo {title} {Picongpu: A fully relativistic particle-in-cell code for a gpu cluster},}\ }\href {\doibase 10.1109/TPS.2010.2064310} {\bibfield  {journal} {\bibinfo  {journal} {IEEE Transactions on Plasma Science}\ }\textbf {\bibinfo {volume} {38}},\ \bibinfo {pages}
  {2831--2839} (\bibinfo {year} {2010})}\BibitemShut {NoStop}%
\bibitem [{\citenamefont {Bart}\ \emph {et~al.}(2017)\citenamefont {Bart}, \citenamefont {Peltz}, \citenamefont {Bigaouette}, \citenamefont {Fennel}, \citenamefont {Brabec},\ and\ \citenamefont {Varin}}]{bart2017}%
  \BibitemOpen
  \bibfield  {author} {\bibinfo {author} {\bibfnamefont {G.}~\bibnamefont {Bart}}, \bibinfo {author} {\bibfnamefont {C.}~\bibnamefont {Peltz}}, \bibinfo {author} {\bibfnamefont {N.}~\bibnamefont {Bigaouette}}, \bibinfo {author} {\bibfnamefont {T.}~\bibnamefont {Fennel}}, \bibinfo {author} {\bibfnamefont {T.}~\bibnamefont {Brabec}}, \ and\ \bibinfo {author} {\bibfnamefont {C.}~\bibnamefont {Varin}},\ }\bibfield  {title} {\enquote {\bibinfo {title} {Massively parallel microscopic particle-in-cell},}\ }\href {\doibase https://doi.org/10.1016/j.cpc.2017.06.004} {\bibfield  {journal} {\bibinfo  {journal} {Computer Physics Communications}\ }\textbf {\bibinfo {volume} {219}},\ \bibinfo {pages} {269--285} (\bibinfo {year} {2017})}\BibitemShut {NoStop}%
\end{thebibliography}%
\end{document}